\documentclass[%
 reprint,
 amsmath,amssymb,
 aps,
]{revtex4-1}

\usepackage{graphicx}
\usepackage{dcolumn}
\usepackage{bm}
\usepackage[normalem]{ulem}
\usepackage{enumerate}

\begin{document}

\title{The Effects of Taxes on Wealth Inequality in Artificial Chemistry Models of Economic Activity}

\author{Wolfgang Banzhaf}

\affiliation{
 Department of Computer Science and Engineering, Michigan State University, East Lansing, MI, USA\\ 
 banzhafw@msu.edu\\
}%


\date{\today}

\begin{abstract}
We consider a number of Artificial Chemistry models for economic activity and what consequences they have for the formation of economic inequality. We are particularly interested in what tax measures are effective in dampening economic inequality. By starting from well-known kinetic exchange models, we examine different scenarios for reducing the tendency of economic activity models to form unequal wealth distribution in equilibrium. 
\end{abstract}

\maketitle

\setcounter{figure}{0}

\section{INTRODUCTION}
\label{sec:Introduction}

Today's societies suffer, for the most part, from a form of economic inequality that seems very difficult to treat or even only to {\it attempt} to remedy.
Governments and economists have made repeated efforts to address this problem, realizing that it is connected to many other serious survival problems of contemporary societies, like climate change \cite{roberts2001global}, loss of biodiversity \cite{mikkelson2007economic}, racial and gender injustice \cite{stolzenberg2006race, seguino2000gender}, health problems \cite{leigh2009health} and others. 

Historians have studied the problem in a perspective across times and generations, and found that it is part of a complex of problems that cause societies to collapse \cite{diamond2005collapse}. Simulation models have been formulated that corroborate the causal relationship between economic inequality and societal collapse \cite{motesharrei2014human}. In turn, some historians have delved into a historical analysis of the reasons for the {\it retreat} of economic inequality, finding societal collapse among them. Their more general findings point to other probably even more harmful and violent causes, catastrophic events like epidemics, wars and revolutions that - together with societal collapse - are virtually the only reasons for a retreat of inequality (see the comprehensive study of Scheidel \cite{scheidel2018great}). Economic inequality seems to be entrenched and potential non-harming  remedies seriously lacking.

Natural scientists have weighed in on this discussion as well, pointing out that situations of equality in both nature and society can be compared and are  ideal symmetrical 
states, bound to disappear into non-symmetrical states as soon as the pressure for equality is reduced \cite{scheffer2017inequality}. 
Studies have found that inequality of ''wealth'' - properly defined - also exists in the animal world \cite{chase2020comparison}, curiously in a similar distribution as among humans. The obvious question to ask then is, whether the unequal distribution in quantities like wealth or species abundance, in other words in {\it stocks} in both natural and social systems, and to a lesser extent the unequal distribution in quantities of flows like income or energy are expressions of universal laws that are acting in both the natural and the human-made (social) world. Econophysics is one of the fields that asks such questions \cite{mantegna1999introduction,sinha2010econophysics} and our contribution here will address some of the assumptions of such models in the sections below. 

Among economists and social scientists there is agreement about the status of economic inequality in societies around the globe, with the United States a particularly egregious example \cite{saez2016wealth,piketty2014capital}. Governments have developed tools to address such issues, with taxes among the most widely used. As is well known, taxes actually play multiple roles in societies, and those roles should be
conceptually kept separate if one wants to apply those tools. The three major roles of taxes are (i) to generate income for government entities on the national, state or local level in order to offer services to the population; (ii) to redistribute wealth between segments of a society; and (iii) to penalize and discourage certain habits or uses among the population. For clarity purposes, here we prefer to discern the usage of this term and speak of {\it government taxes and fees} for purpose (i), {\it redistributive taxes} for purpose (ii) and {\it penalty taxes} for purpose (iii).  

Among what is the general (and less discerning) usage of the term ''tax'', a number of different tools have been applied to generate cost-recovery for government services, like sales taxes, value added taxes, income taxes, wealth taxes, inheritance taxes, luxury taxes, etc. 
Easiest to administer are certainly those taxes that are applied during the event of a transaction, like a sales tax at the moment when an actual sale happens.
Less easy to administer are taxes that span a period of time, like taxes for earnings or interest, to be collected when a transfer of earning
or interest happens periodically. Both of these taxes can be tied to relatively easily measurable flows. The most difficult taxes to administer are those that are tied to a stock, like wealth or inheritance taxes. Disputes quickly ensue about the measurability and comparability of stocks, their discountability etc, and, depending on the interest of a party, minimization strategies are applied to circumvent the tax duty following from a tax on stocks of any kind. An answer to such behavioural variations is probably offered by game theory, but for the moment, we shall have to postpone consideration of tax avoidance strategies and focus on the results of applying idealized taxing policies. Note also that we do not consider any other method for avoiding economic inequality in this paper, though there might be other routes to improve equality like the creation or fostering of a gift economy which, however, would require basic systemic changes to economic activity.

Among economists, there is an ongoing discussion about the relation between taxation and productivity. Under the assumption that productivity and taxation are conflicting goals in a society, optimization methods seem to be an appropriate approach to reconcile those conflicting goals. While economists formulated and studied optimal taxation theories \cite{sandmo1976optimal,heady1993optimal,kocherlakota2006advances}, progress in techniques like computer simulation has allowed a whole set of models to be explored in this arena. The idea of such models is that while they will often be very simple abstracted models of reality, their results can teach us a lot about general taxation effects which can then be refined and underpinned by theoretical investigations. The econophysics models of wealth formation as well as the study of phenomena equivalent to wealth in ecologies are two examples already mentioned earlier. 

Here we suggest another route to explore taxation models, by virtue of very simple multi-agent systems called Artificial Chemistries \cite{banzhaf2015artificial,dittrich2001artificial}. Artificial chemistry models are based on an analogy between the system under consideration and a chemical reaction system. Stochastic interactions (''reactions'') happen between agents called ''molecules'' which obey predetermined rules. In contrast to many other MAS, however, the behavioural rules of these molecular agents are simple and identical among agents of the same type, thus allowing the collective effects to emerge clearly from the set of chosen rules.
 
Here we are interested exclusively in distributive tax effects. But before we go into details of such AC models of wealth distribution, the major result of the current investigation can be summarized as follows: These very simple abstracted models of (economic) interaction between agents result in a general outcome: income tax is overrated as a means to achieve economic equality, or a remedy to at least reduce economic inequality. Even income taxes at the high end of the scale (progressive, high marginal rates) do only slightly dampen, but do not eliminate economic inequality. One needs to introduce a wealth tax to tackle the problem. Due to the simplicity of the model, this result is general and can be expected to hold even if more complicated interactions or structures are envisioned. 
We do not rule out, of course, that from a practical point of view, a combination of different taxes needs to be employed, but a proper wealth tax needs to be the major component of any system that hopes to substantially reduce economic inequality. 

The rest of this paper is organized as follows: The next section (Section \ref{sec:distribution})  briefly reflects the current knowledge on income and wealth distribution, Section \ref{sec:kinetic} then discusses the kinetic exchange model of econophysics and its key assumptions and results. Section \ref{sec:OurABM} explains our artificial chemistry model of economic activity in detail. Section \ref{sec:results} then provides the simulation results of the model using different scenarios with and without different types of taxes. 
Section \ref{sec:discussion} discusses various counter-arguments against the primary result and some practical implementation issues, Section \ref{sec:conclusion} concludes by putting the results in perspective. An Appendix gives more details on distributive effects of some of the models studied.\\

\section{INCOME AND WEALTH DISTRIBUTION} \label{sec:distribution}

\begin{figure}
\centerline{\includegraphics[width=0.3\textwidth]{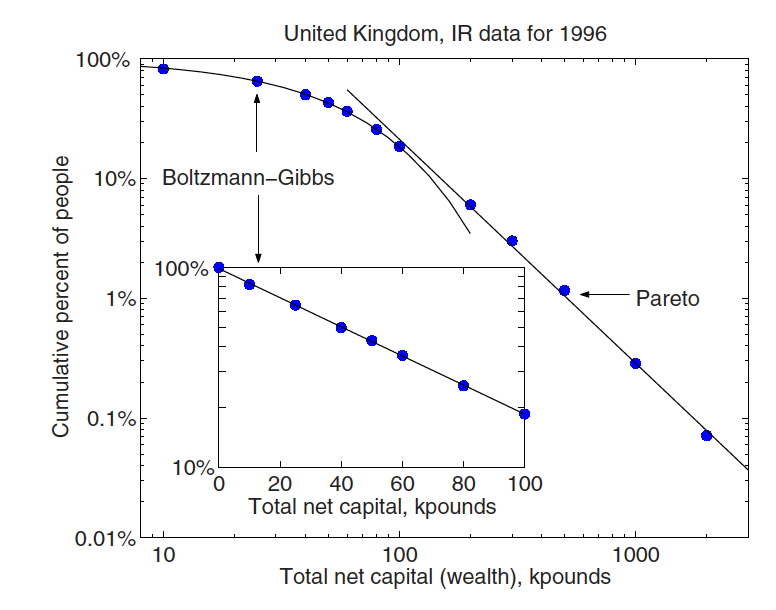}}
\vspace{5mm}
\centerline{\hspace{3mm} \includegraphics[width=0.3\textwidth]{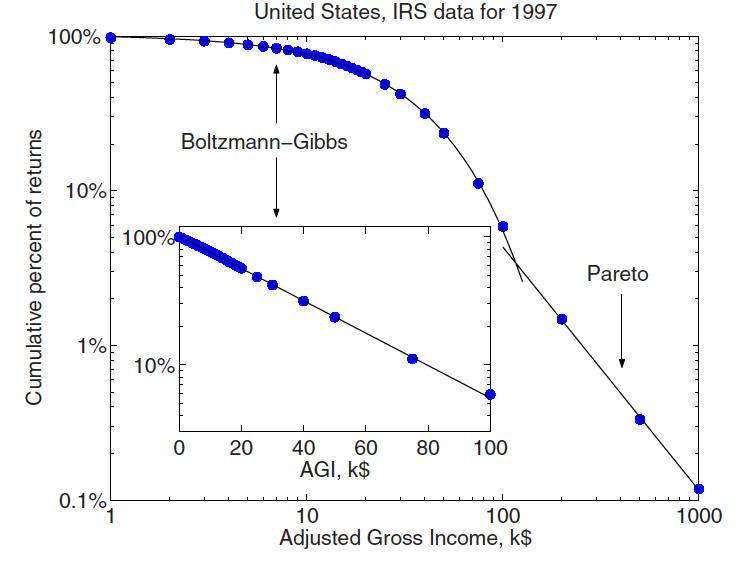}}
\caption{Wealth distribution for 1996 (UK), top, reproduced from \cite{druagulescu2001exponential}, and income distribution for 1997 (US), bottom, reproduced from \cite{dragulescu2000statistical}. }	
\label{fig:distributions}
\end{figure}

Income and wealth distribution in virtually all countries of the world cannot be described other than unequal. Unfortunately, this is not a temporary situation brought about by some economic downturn or some unsuccessful or incapable attempts of governments on behalf of their citizens. Instead, it is a systemic, long-term problem of virtually all present and historic societies, the source of many of the problems that caused and still cause the collapse of their organization and of the well-being of their citizenry. As Piketty has pointed out in his large-scale study \cite{piketty2014capital} this problem will continue in the 21st century until we find a means to seriously fight its root causes. 

The history of human economies provides a rich field for learning about trends and tendencies. The anthropology/archaeology pair Kohler and Smith \cite{kohler2018} provide a larger overview of the history of inequality in human societies. The Italian economist and sociologist Pareto studied wealth distribution in Europe already in the 19th century. 
He found that wealth distribution for the richer segments of a society follows a power law, today known as the Pareto law \cite{pareto1897}. This sector refers to the upper echelons of wealth and 
income, whereas the lower part of the distribution curve can be fit well with an exponential or Gibbs distribution or a log-normal distribution. Yakovenko and Rosser \cite{yakovenko2009colloquium}, based on \cite{dragulescu2003statistical} offer a good review of data and models of income and wealth distribution. However, data on wealth are difficult to come by and often proxies have to be used. Figures from the UK on wealth 
(derived from inheritance) for 1996 and from the US on income distribution for 1997 produce a similar picture (see Figure \ref{fig:distributions}), though. 

Commenting on income distribution, Chakrobarti et al write \cite{chakra2013}:
\begin{quote}
''These observed regularities in income distribution may thus indicate a 'natural' law in economics.''
\end{quote}

\begin{figure}
\centerline{\includegraphics[width=0.46\textwidth]{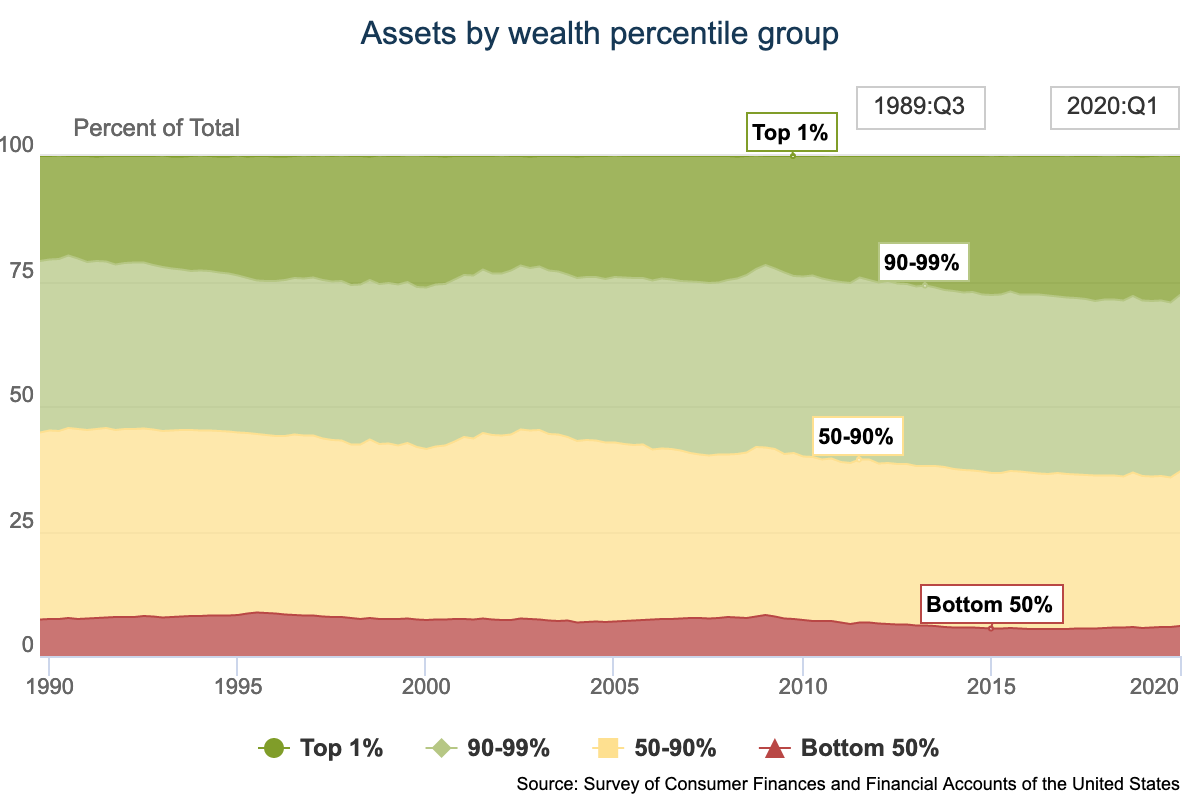}}
\caption{Wealth distribution for top 1\%, 90-99\%, 50-90\% and bottom 50\% of population, development 1989-2020. \cite{fed2020}}
\label{fig:fed-chart}
\end{figure}

For the United States, further evidence comes from statistics showing the development of wealth distribution across percentiles of the population, Figure \ref{fig:fed-chart}. As of early 2020, the top 1\% own more than 25\% of the assets, while the bottom 50\% have a
share of approximately 5-7\% of the assets. 

Again, given that wealth seems to follow the same distribution as income, it makes sense to use a universal explanation for these
tendencies, even if details might be different. 

\section{THE KINETIC EXCHANGE MODEL} \label{sec:kinetic}

The kinetic exchange model for economic activity starts with the hypothesis that the economic exchange activity of individual 
economic agents in the form of trades can be compared to the movement and encounter of gas particles exchanging energy \cite{dragulescu2000statistical,chakraborti2000statistical}.
In place of entropy maximization in the case of energy exchange they would follow utility maximization principles in the case of
economic trades. Money would take the place of energy. Benoit Mandelbrot \cite{mandelbrot1960} has succinctly summarized this idea: 

\begin{quote}
''There is a great
temptation to consider the exchanges of money which occur in economic interaction as analogous to the exchanges of enegy which occur in physical shocks between molecules. In the loosest possible terms, both kinds of interactions ''should'' lead to ''similar'
states of equilibrium. That is, one ''should'' be able to explain the law of income distribution by a model similar to that used in statistical thermodynamics ...''
\end{quote}

Mandelbrot goes on to point out that actual income distribution is different, and considers models to accommodate that difference. 
But the general idea of these models is that if you have a population of agents that interact in a random fashion with each other, the
distribution of income and wealth approach certain equilibrium distributions well known from physics.

Key assumptions of the simplest of these models of income and - by way of transfer - wealth distributions are (i) a closed economic system in which the number of economic agents and the total wealth of the system remain constant, i.e. that trading exchange is the prevalent mode of economic activity; (ii) trading is restricted to two-agent  interactions; (iii) there is no negative wealth, thus an exchange cannot result in such and debt needs not be considered; (iv) the exchange is symmetrical, in that the basis of the exchange is normally a fixed amount or a percentage of the sum of the wealth of the participating agents; (v) the exchange conserves money, i.e. the amount given to one agent is taken from the other so that the exchange does not change the total amount of wealth in the system; (vi) an exchange process does not depend on previous exchange processes, so the dynamics is Markovian. 

Key results of these models are that the income/wealth distribution develops indeed unequally among the population of agents, and, regardless of what the initial conditions of the exchange dynamics are, tends to an equilibrium that has a distribution that looks similar to an exponential (Gibbs) or log-normal (Gibrat) distribution.  The details vary based on model assumptions. For example, if there is an additional assumption of savings (reserved amounts of wealth not available for the economic exchange) then such models tend to reflect wealth distribution in societies more closely. In particular, the tail of the previous distribution is now modified and follows a power law (Pareto law), just as empirically observed. Details can be found in \cite{yakovenko2009colloquium,chakra2013}.

The kinetic exchange model has been criticized by economists as being unrealistic and even, to a degree, misleading \cite{gallegati2006worrying}. One key criticism is 
\begin{quote}
"The industrialised economies of the West, and increasingly of Asia, are emphatically not a conservative
system: income is not, like energy in physics, conserved by economic processes. Therefore, it is a fundamental
fallacy to base economic models on a principle of conservation. Yet this is an inevitable consequence of
exchange-only models, since exchange {\it is} a conservative process.''
\end{quote}
We don't think that this criticism is indeed the core of the problem because it is not a fundamental limitation of exchange models. 
There are other kinetic exchange models that allow for some of the
money to go into taxation \cite{guala2009,bisi2009,duering2008}, thus reducing the amount that is available to the agents subsequently. 
We show in this contribution how easily such models can be formulated. 
In the same vain one could imagine that money is gained in the process of exchange, thus allowing both agents to benefit 
from the exchange process and generating a growing economy \cite{smith1776,ricardo1817}. 

However, the issue is probably related to 
the fact that kinetic model are often theoretical and "... much of the econophysics community appears to think that simply doing good science is sufficient to have the work recognised, rather than relating to the motivations and incentives of policy makers ...'' \cite{ormerod2016}. In the following, we intend to discuss an example of an easily formulated model with an exclusive   
focus on the purpose of the model: to understand distributional effects. 

\section{Using an Artificial Chemistry as an Agent-based Model for Economic Activity} \label{sec:OurABM}

Our agent-based model is based on an Artificial Chemistry (AC) \cite{banzhaf2015artificial} with the idea that it is the {\it interaction} of agents that is the most important driver of {\it distribution} of wealth and income. The detailed nature of that interaction will determine the exact distribution of wealth in a population of agents, but for our modelling purposes here, we study as simple a model as possible. 

\begin{figure}
\includegraphics[width=0.3\textwidth]{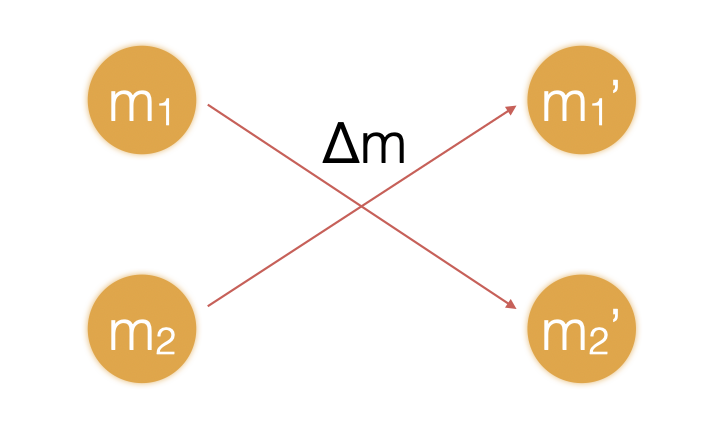}
\caption{Agents with wealth $m_1, m_2$ at time $t$ interact and exchange money $\Delta m$ in the process, resulting in agents with 
wealth $m_1', m_2'$ at time $t+1$.}	
\label{fig:chem_react}
\end{figure}

The baseline model assumes that all agents are homogeneous in their behaviour, while possibly possessing different amounts of wealth. 
Their interaction is based on the following AC rules. In each iteration randomly chosen agents $i$ and $j$, possessing amounts of respective wealth of $m_i$ and $m_j$ encounter each other and exchange a good for an amount of money $\Delta m$ (see Figure \ref{fig:chem_react}). 
In terms of an AC, the population is in a well-mixed reaction vessel without inflow or outflow.
We are not interested in studying the flow of goods in the economy, only the flow of money. Thus, we assume something of value will flow in the counter direction of the money flow, without further specifying the nature of that flow. We are not allowing negative wealth, so $\Delta m$ will be determined as a random percentage of the {\it smaller} of the amounts $m_i$ and $m_j$. Alternatively, we could define the attempt to exchange the larger amount of money as an ''elastic collision'' of the agents, but that would simply force us to draw another pair of agents, and therefore delay the relaxation of the system. 

Suppose $m_i < m_j$, and a percentage $p$ flows in the direction $m_j$. Then each agent updates their ''wealth'' in iteration $t+1$ with the following equations:
\begin{equation}
\begin{aligned}
m_i (t+1) = m_i (t) - \Delta m \\
m_j (t+1) = m_j (t) + \Delta m
\end{aligned}
\end{equation}
with 
\begin{equation}
\Delta m = p \cdot m_i
\end{equation}
Note that only one agent, randomly chosen, receives money, the other is supposed to receive a corresponding good. The total amount of money is, of course, unchanged by this operation, but the operation is asymmetrical. Since the smaller amount determines the actual flow, if one of the agents is getting very poor with a wealth close to $0$, the amount flowing is getting correspondingly smaller, and the  
exchange creeps to a halt. We do not modulate the probability of exchanges based on their size in this baseline model.

As has been pointed out above, the kinetic exchange models of econophysics assume a symmetrical exchange, based on the sum of both agents' wealth, and having money flowing in both directions. This makes sense when thinking about energy flows in a physical system, but here we are discussing a different system where the assumption of symmetry is not justified. Artificial chemistries are no stranger to asymmetrical processes, in fact, it is their general case.

The result of the kinetic exchange model is that it equilibrizes much faster and to a less unequal distribution. {\it Our} model's equilibrium state is extreme, with one agent of the population holding all the wealth, and all others holding none, even when starting from a fully equal 
wealth distribution at the outset. The relaxation times for approaching this equilibrium are, however, much longer, so that in a reasonable simulation it is never reached. 

\begin{figure}
\includegraphics[width=0.23\textwidth]{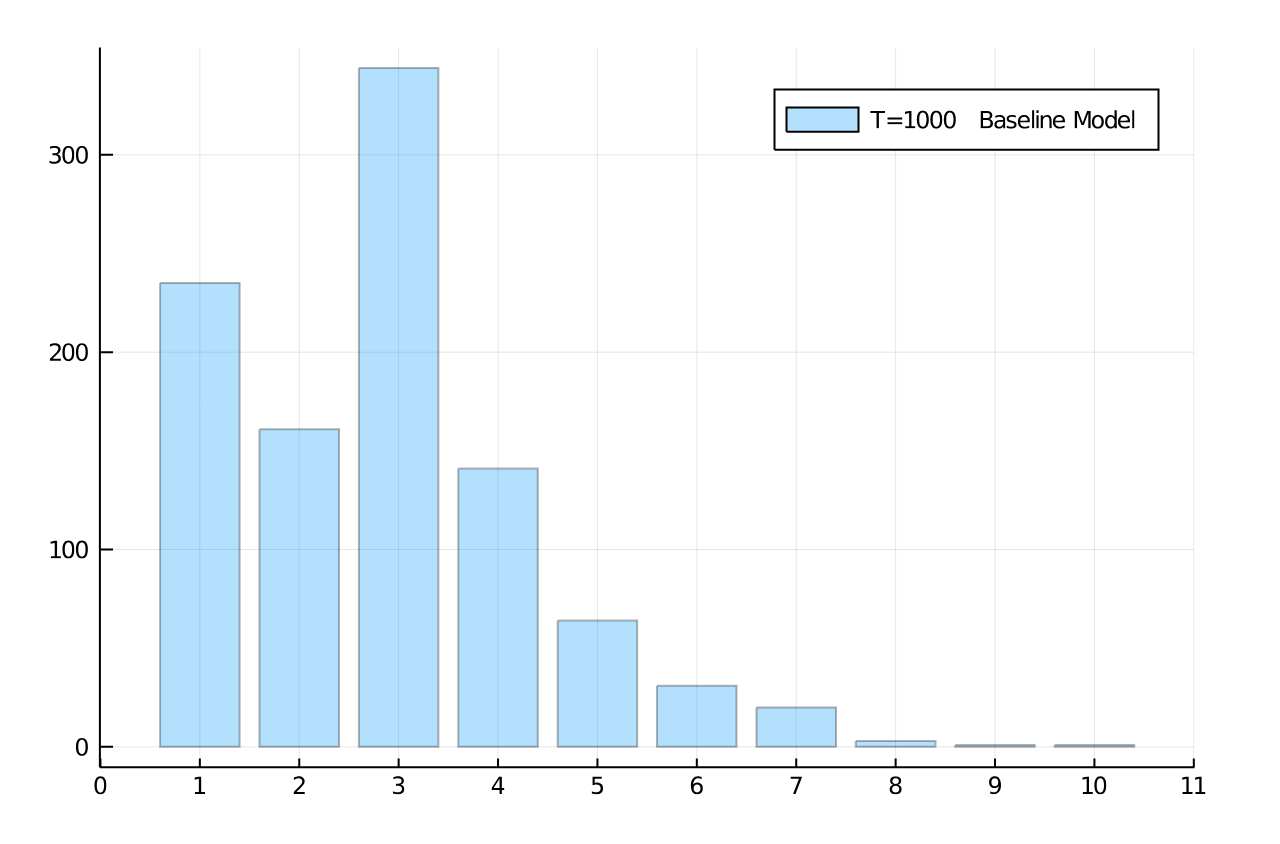}
\includegraphics[width=0.23\textwidth]{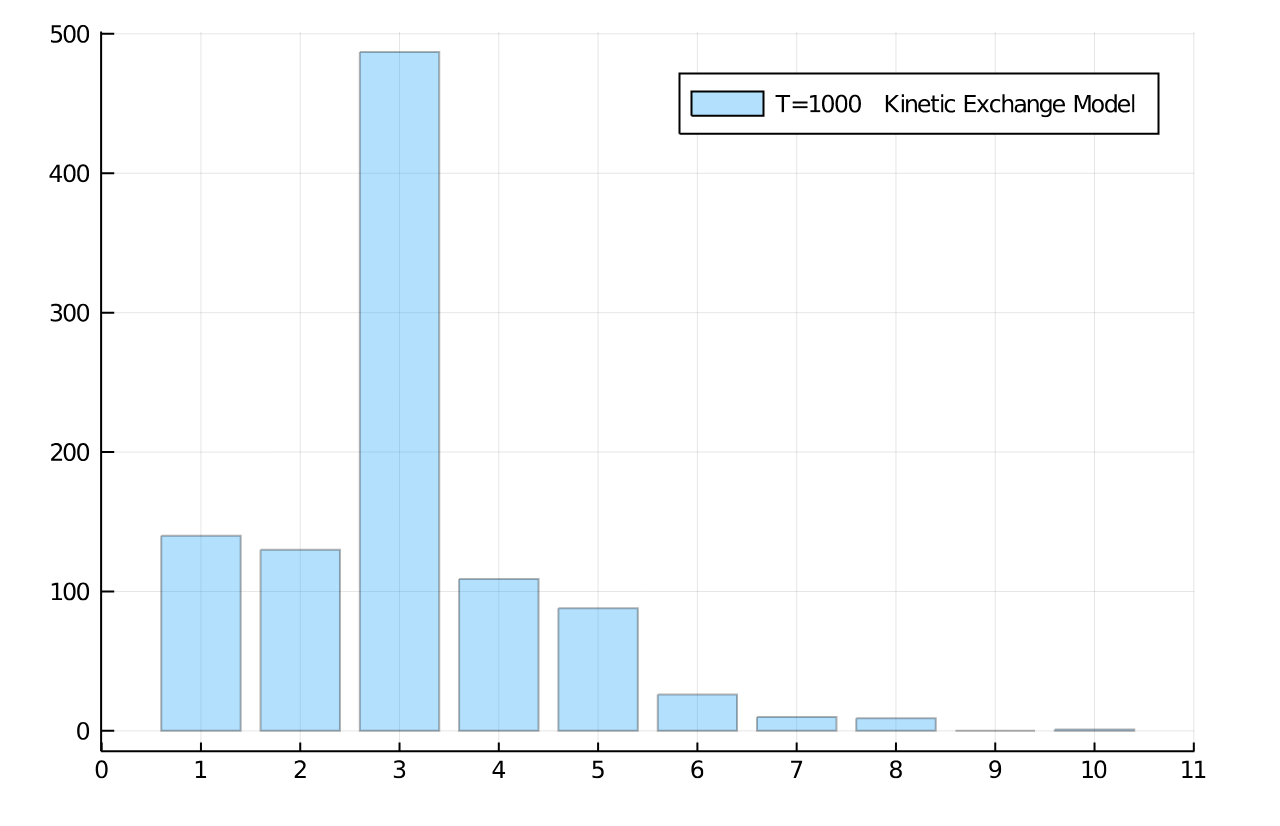}
\includegraphics[width=0.23\textwidth]{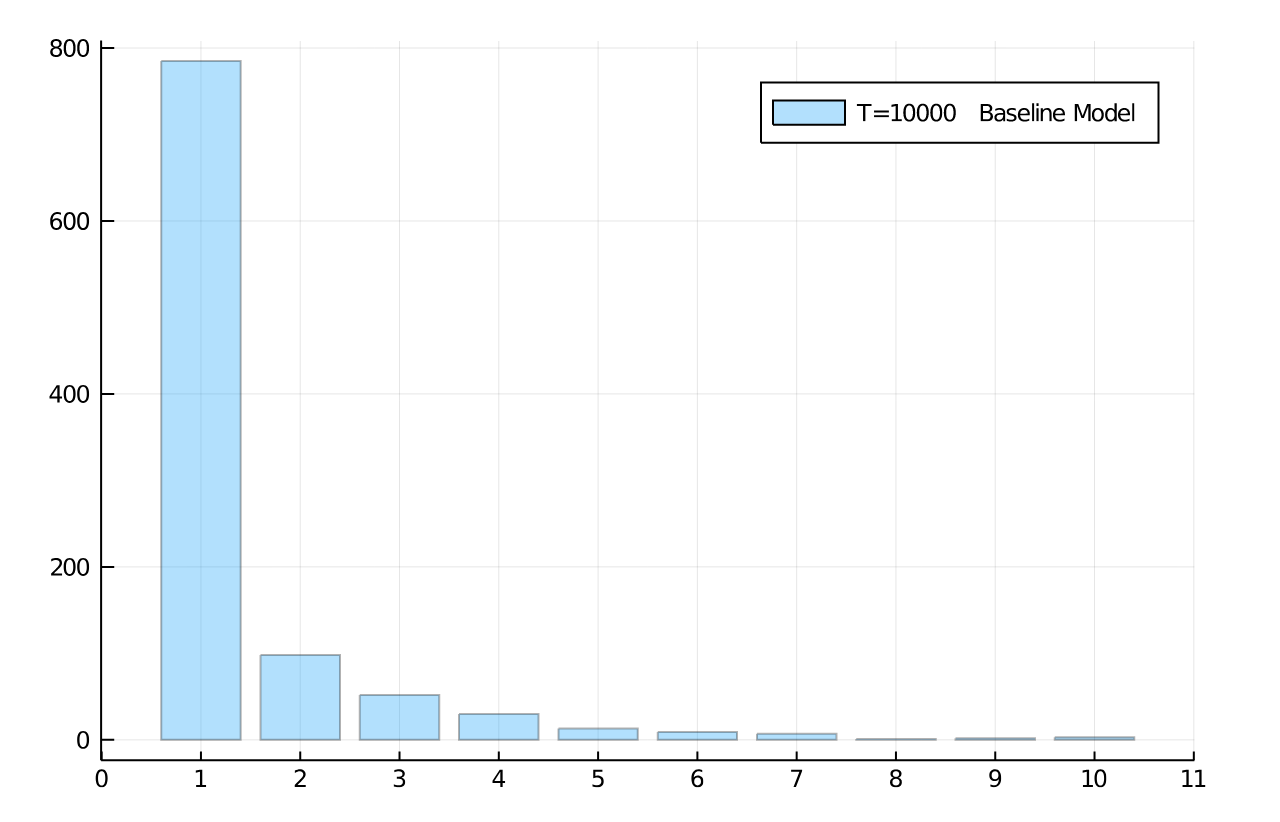}
\includegraphics[width=0.23\textwidth]{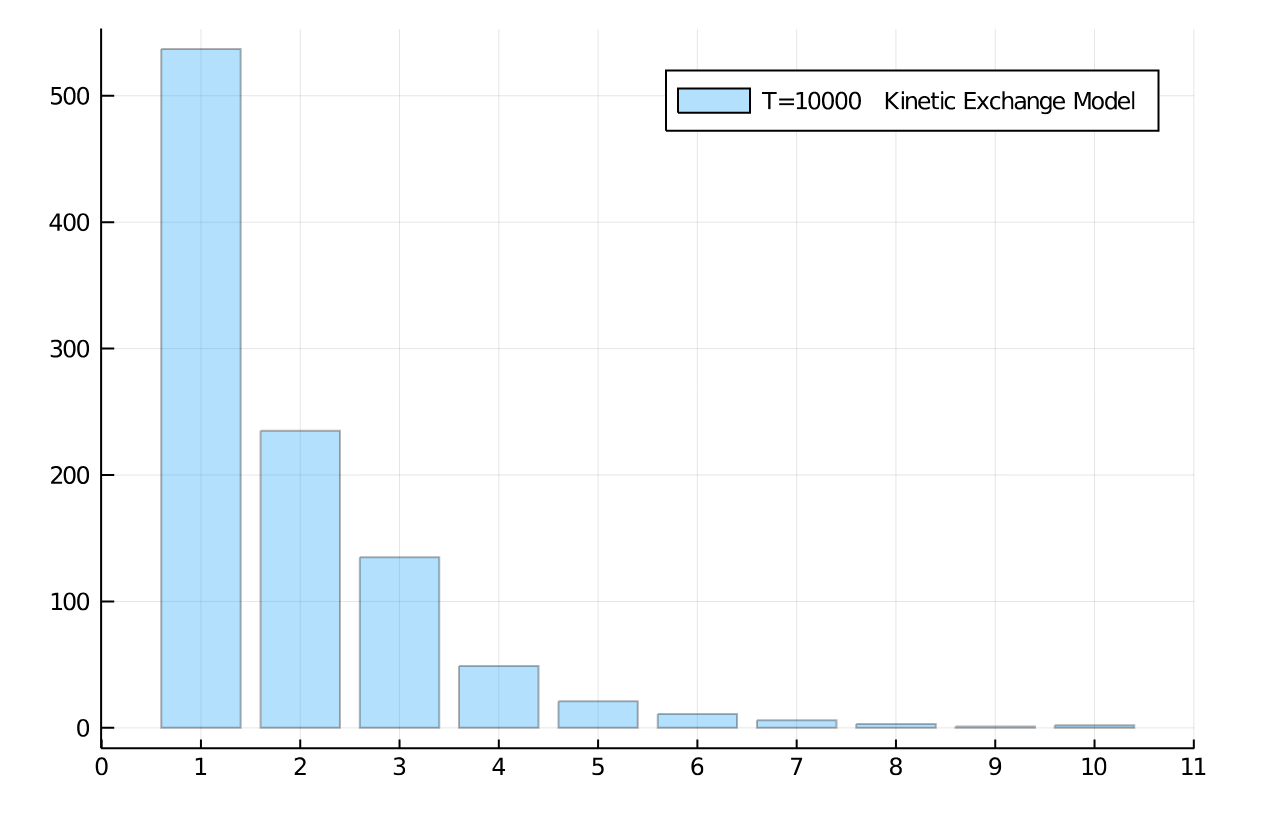}
\includegraphics[width=0.23\textwidth]{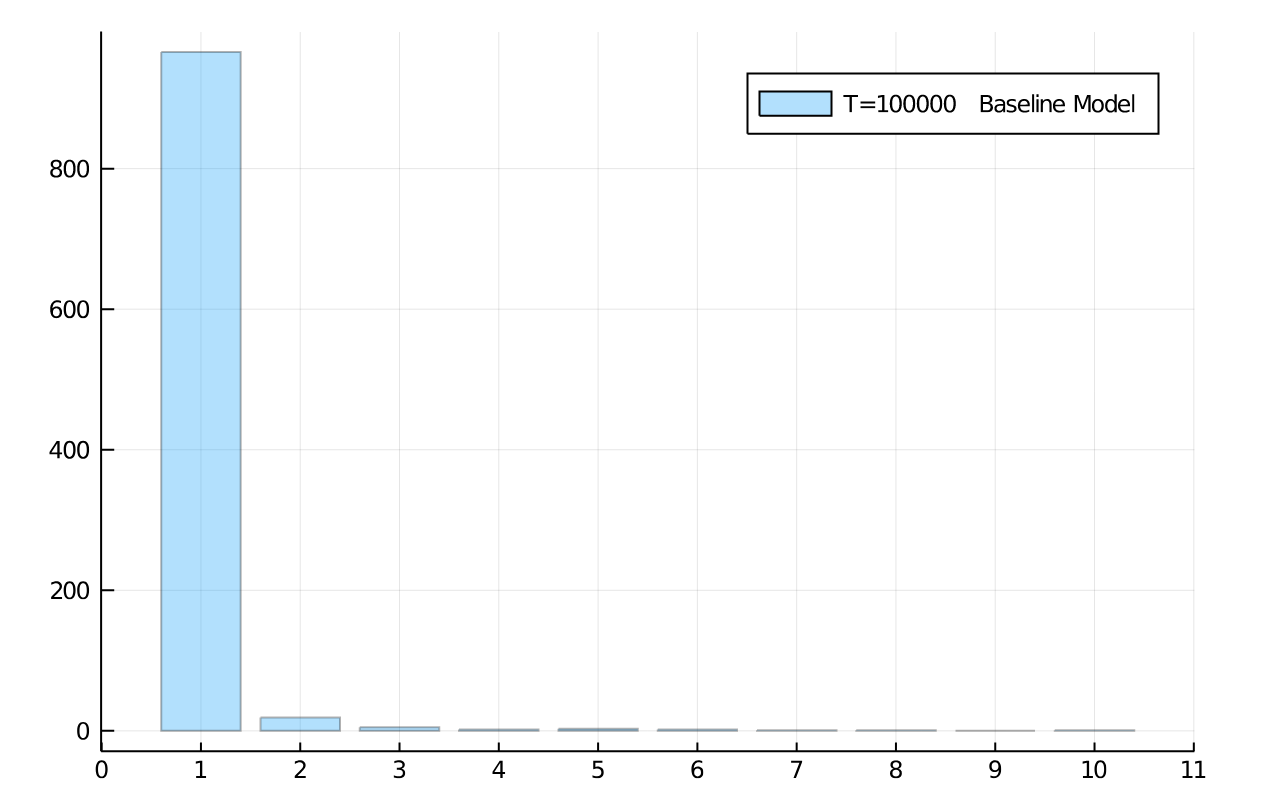}
\includegraphics[width=0.23\textwidth]{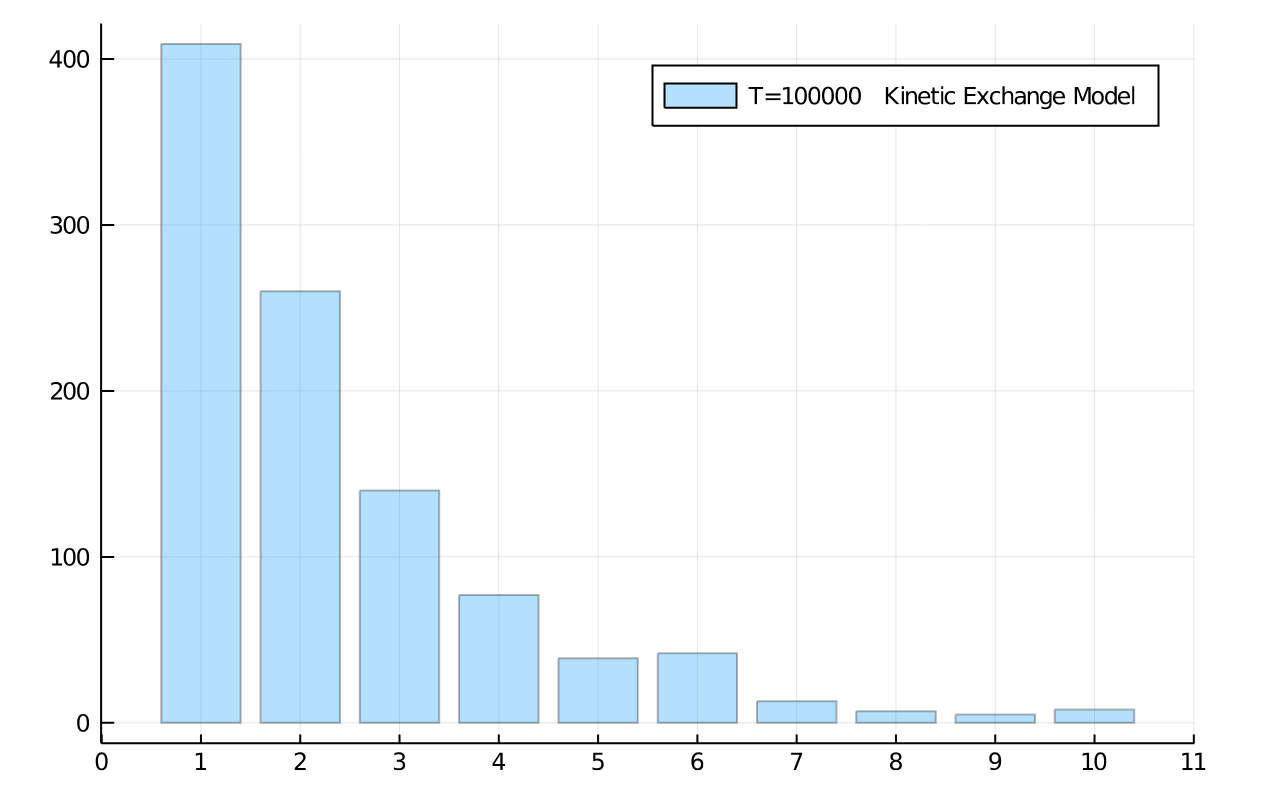}
\caption{Wealth distribution in population of 1,000 agents (10 bins, a.u.). Comparison of baseline economic exchange model (left) and kinetic exchange model (right) at $t=1,000; 10,000;$ and $100,000$. Initial distribution was equal among all 1,000 agents.}	
\label{fig:compare}
\end{figure}

Figure \ref{fig:compare} compares our AC model with the kinetic exchange model of Dragulescu and Yakovenko \cite{dragulescu2000statistical} at different iterations. That model determines the exchange of money as a percentage of the
average wealth of the two agents: 
\begin{equation}
\Delta m = p \cdot (m_i + m_j)/2
\end{equation}

We can see clearly that the inequality resulting from the asymmetrical agent model is even higher and forming faster than
the unequal wealth distribution from the kinetic exchange model. Note that both simulations started from an equal distribution of
wealth among all 1,000 agents. But regardless of initial state, both models develop heavy inequality. Many other model variants can be formulated that lead to qualitatively similar results: Even from an ideal state of equal distribution among agents, wealth inequality develops as if it were a natural law. 

\begin{figure}
\includegraphics[width=0.46\textwidth]{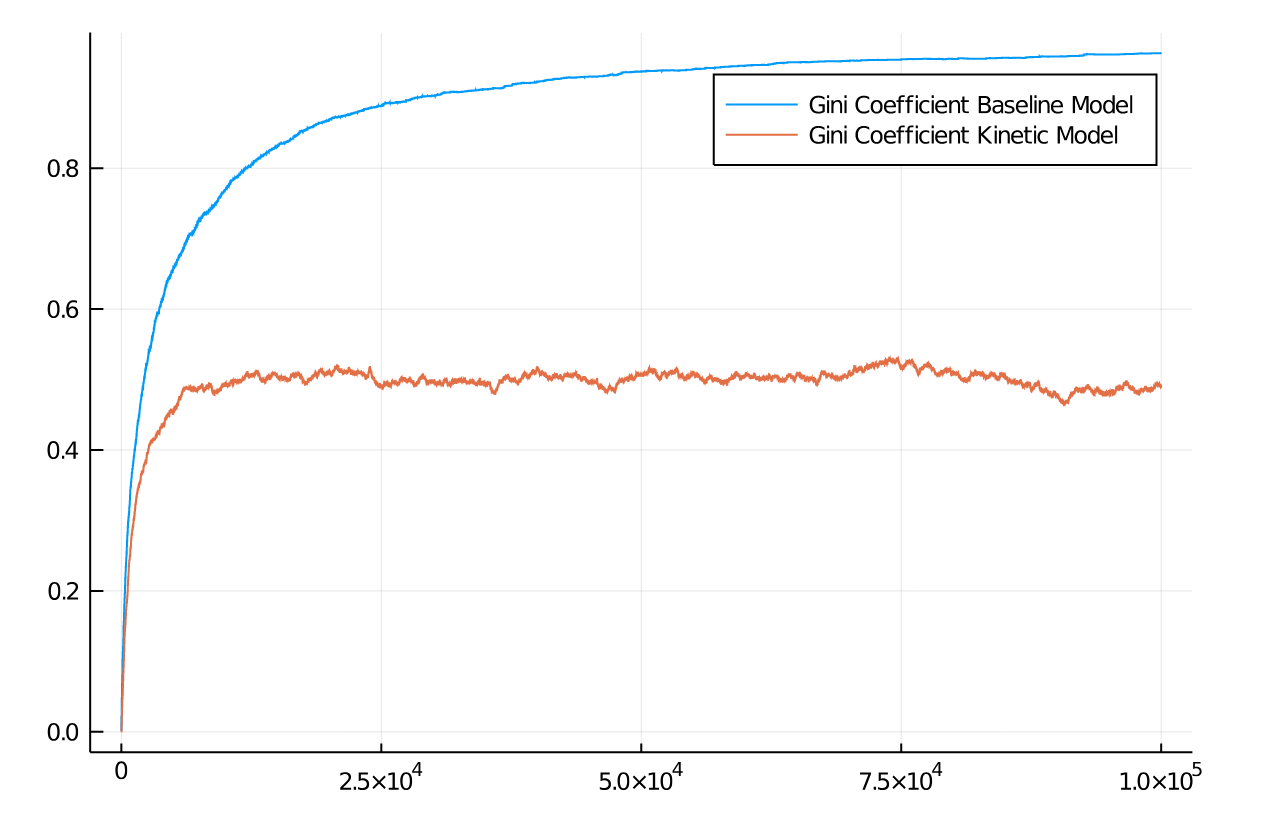}
\caption{Gini coefficients in population of 1,000 agents for the baseline economic and the kinetic exchange model. Initial distribution was equal among all 1,000 agents.}	
\label{fig:gini}
\end{figure}

A classical measure of economic inequality is provided by the Gini coefficient g, $0 \le g \le 1$, defined as:
\begin{equation}
g = \frac{1}{2 N^2 \bar{m}} \sum_{i,j} \| m_i - m_j \|
\end{equation}
Figure \ref{fig:gini} shows the Gini coefficient for both simulations and how it develops over time. As expected, for the kinetic model approaching an exponential distribution, the Gini coefficient fluctuates around $1/2$ which signals a substantial inequality in wealth distribution among agents. For our baseline model, the Gini approaches $1$, the most unequal Gini coefficient possible. Generally speaking, Gini coefficients below $0.25$ indicate high equality in a distribution.

\section{Results} \label{sec:results}

In this section we shall discuss our simulation results on the following different scenarios:
\begin{enumerate}
\item A flat income tax with regular (30\%), high (60\%) and low (5\%) values
\item A progressive income tax between 30\% and 75\%
\item A wealth tax of medium, high and low value
\end{enumerate}
For some of these tax regimes, we also study different redistribution cases.
\begin{enumerate}[I]
\item Redistribution to all tax payers
\item Redistribution to a select group of tax payers (those with negative income, or lower half of wealth distribution)
\end{enumerate}
As emphasized earlier, in these simulations we do not consider tax as income for local, state or federal governments. This is an important, but different consideration for tax usage. Rather, we are interested only in the distributional effects of very simple interaction rules. When we show distributional effects, we also do not show absolute values, we rather bin into deciles of wealth, i.e. agents relative to each other.

\subsection{A Flat Income Tax} \label{sec:fit}

In the flat income tax regime we apply a given tax rate $r$ to the income of each agent, as it develops over periods of time. In the current simulations, we apply the tax every 10 iterations to the difference of wealth an agent has accumulated over this period. Thus, we adjust each agent $i$'s wealth by the following formula:
\begin{equation}
m_i (t) = m_i (t) - r  [ m_i(t) - m_i(t-10) ]
\end{equation}
provided its earnings over the period, $m_i(t) - m_i(t-10)$, are positive. Otherwise, we do nothing. 

Once tax from all agents has been collected in the period, resulting in an amount $T(t)$, we distribute it back to the agents according to (I) equal redistribution; or (II) select redistribution. This updates the wealth of agent $i$ again to:
\begin{equation}
\begin{aligned}
m_i (t) = m_i (t) + \frac{T(t)}{N}  \hspace{2mm} \forall i \in \cal{N} \hspace{5mm} \mbox{(Case I)} \\
m_i (t) = m_i (t) + \frac{T(t)}{| \mathcal{N} _n |}  \hspace{2mm} \forall i \in \mathcal{N} _n \hspace{5mm}  \mbox{(Case II)} \\
\end{aligned}
\end{equation}
where $\mathcal{N} _n$ is the set of all those who loose in the transactions, i.e. those agents with negative income over the period $m_i(t) - m_i(t-10) < 0$.

\begin{figure}
\includegraphics[width=0.23\textwidth]{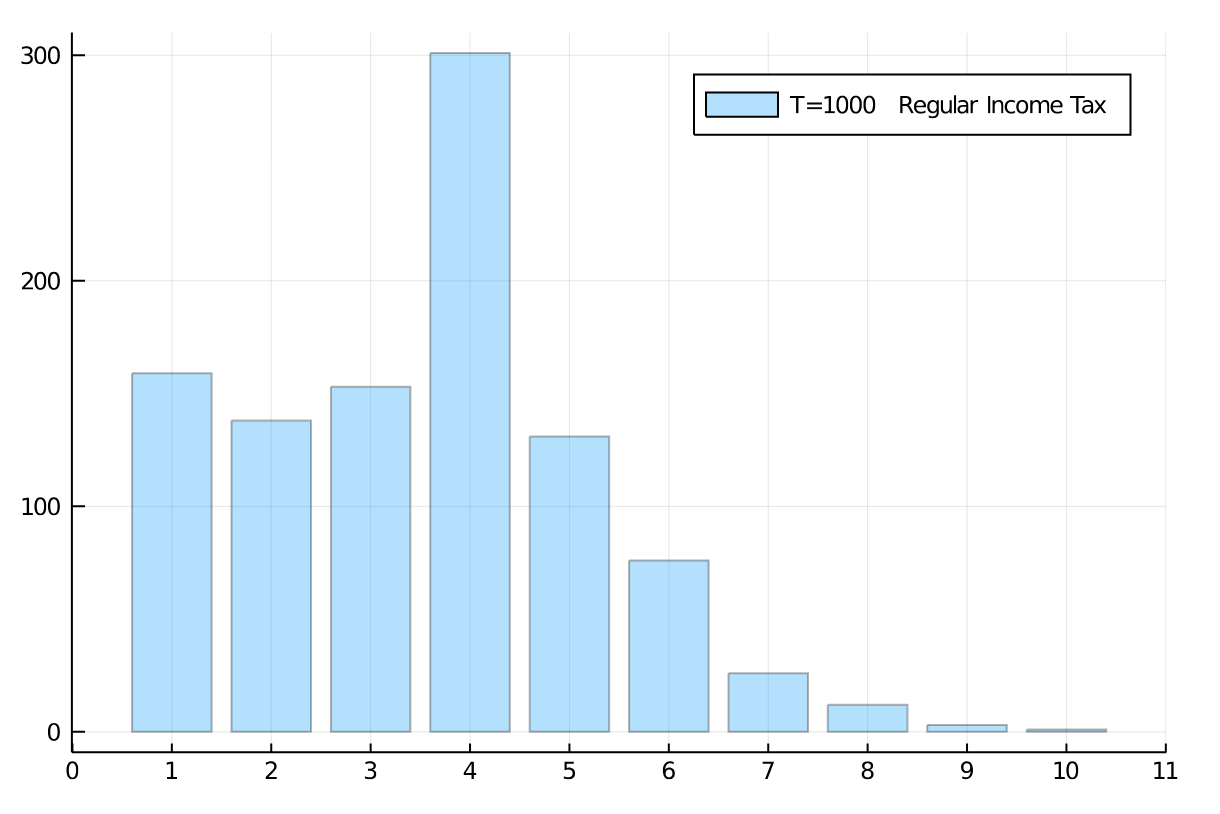}
\includegraphics[width=0.23\textwidth]{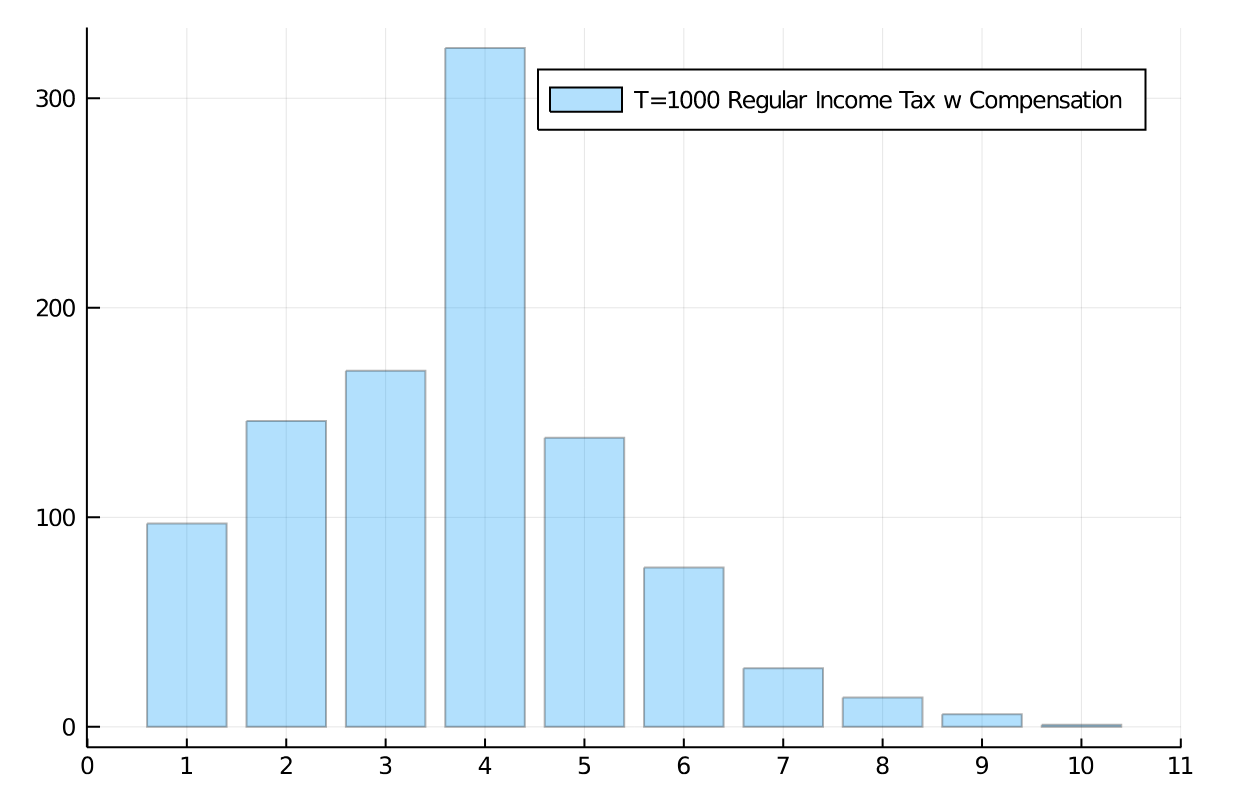}
\includegraphics[width=0.23\textwidth]{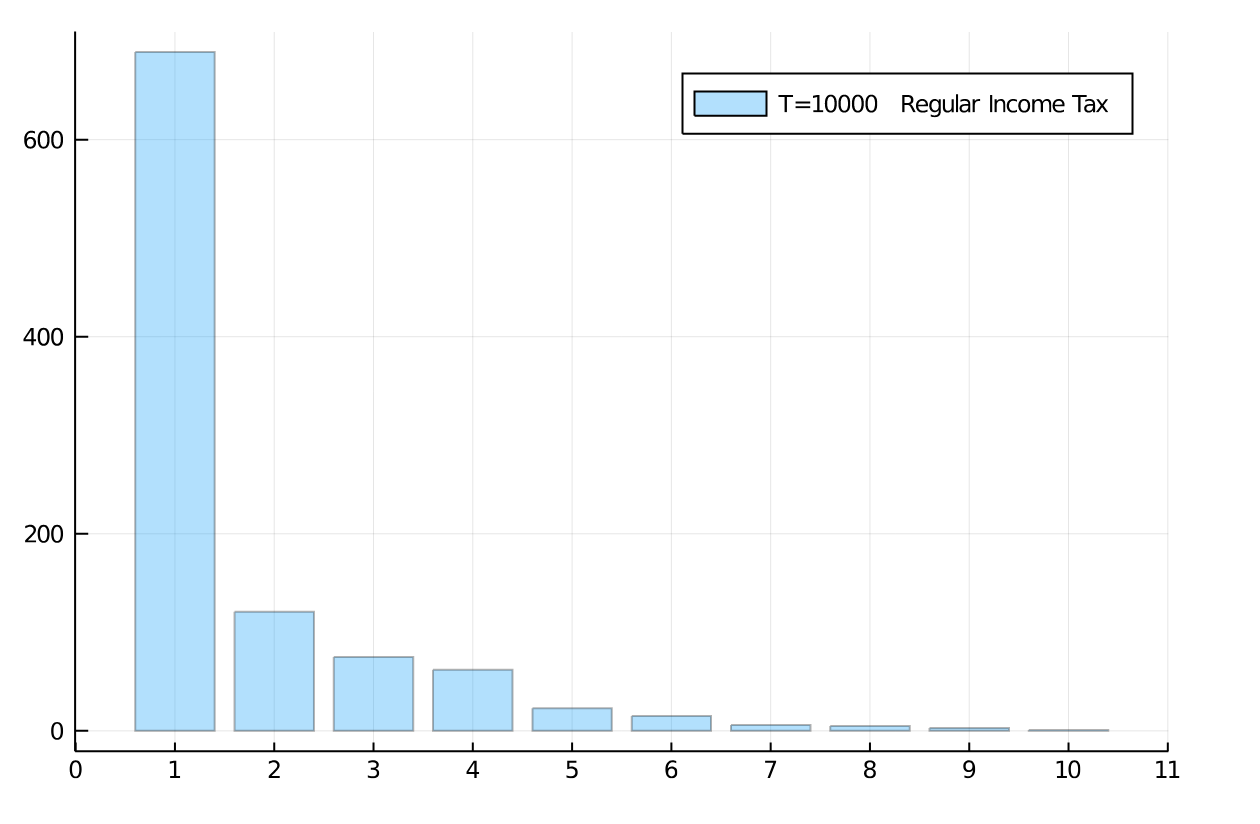}
\includegraphics[width=0.23\textwidth]{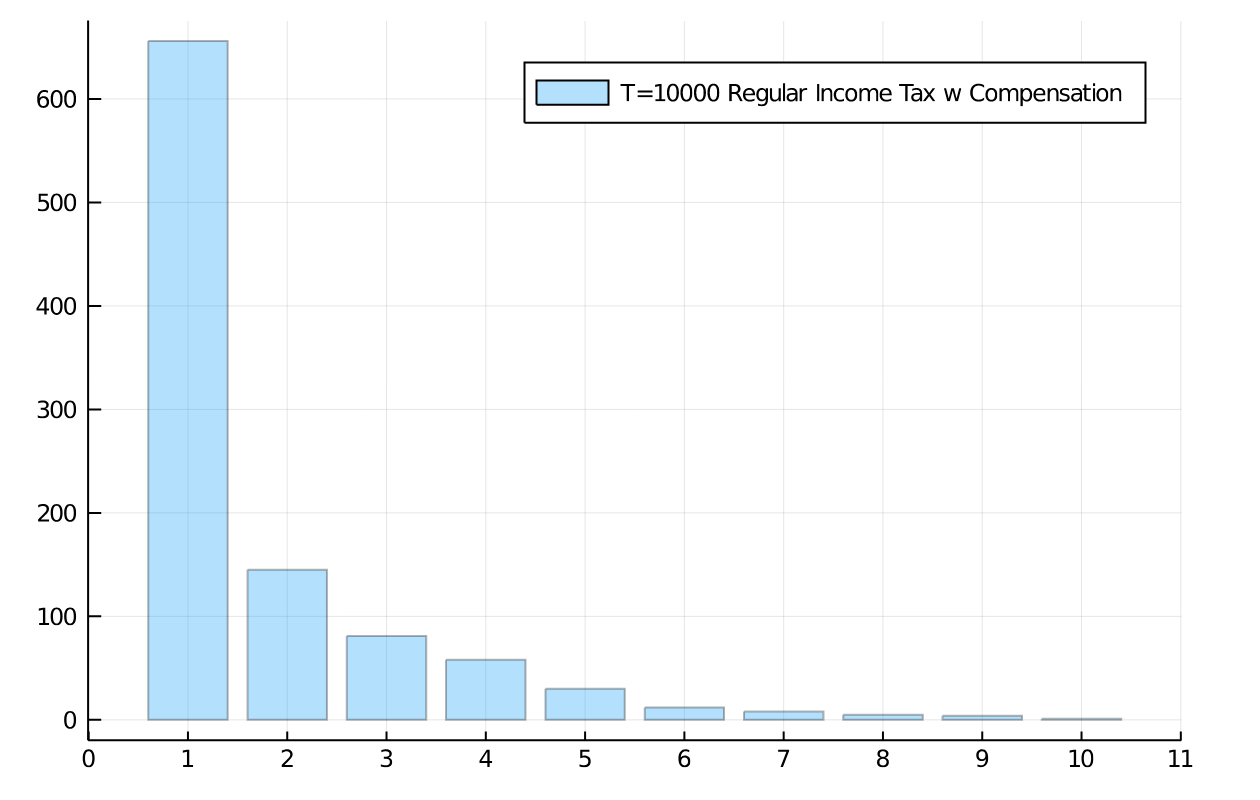}
\includegraphics[width=0.23\textwidth]{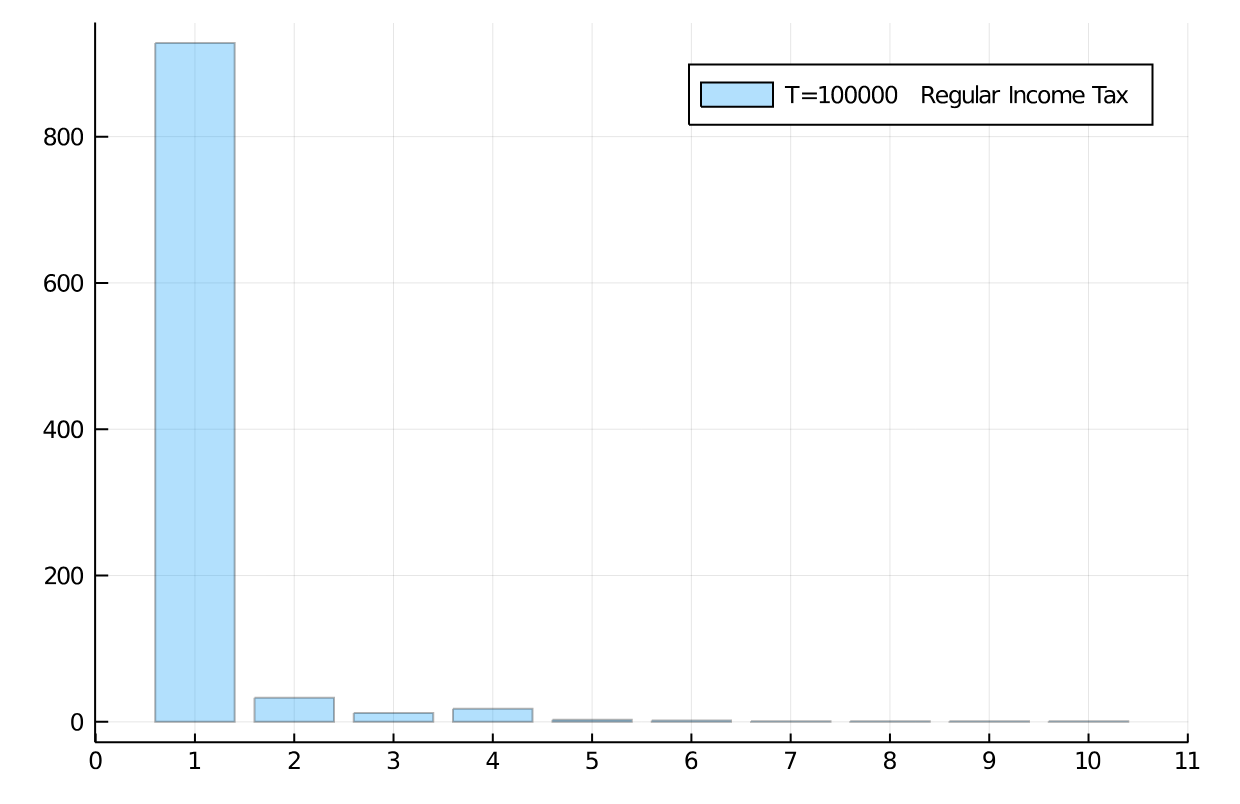}
\includegraphics[width=0.23\textwidth]{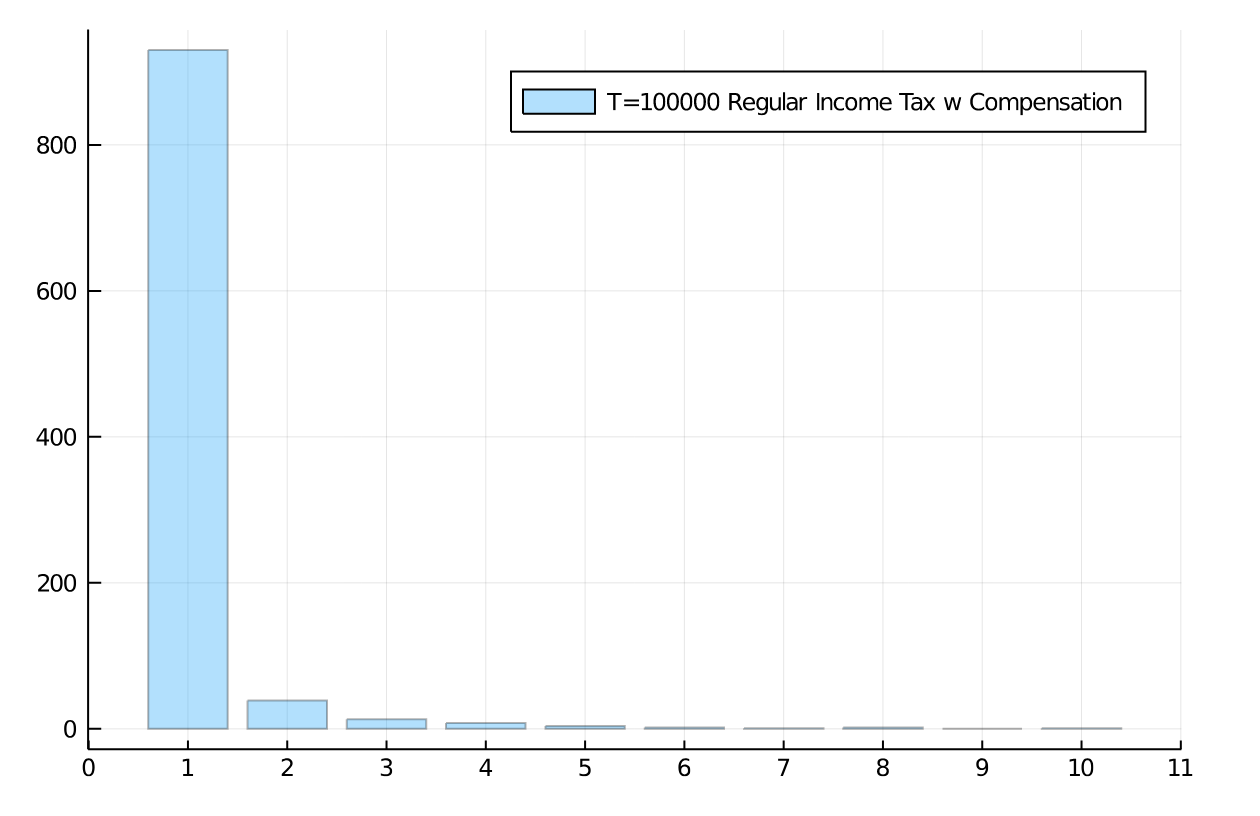}
\caption{Wealth distribution in population of 1,000 agents (10 bins, a.u.). Flat income tax regime, with regular flat tax of 
30\%. Comparison of case l (left) and case II (right) for redistribution policy at $t=1,000; 10,000;$ and $100,000$. Initial distribution was equal among all 1,000 agents. Difference is virtually not visible.}	
\label{fig:flat_income_tax-regular}
\end{figure}

Figure \ref{fig:flat_income_tax-regular} shows the results of applying a flat income tax of $r=30\%$ and subsequent equal redistribution of the resulting taxes to all (case I) or the agents that have lost income in the transaction (case II). While in the early stages of the interactions there is some difference visible, that difference seems to become smaller and smaller as the number of interactions increases. We show the Gini coefficients for these flat tax experiments in the Appendix. The vast majority of agents has become very poor, despite the application of an income tax, and despite starting out with an exact equal distribution of wealth.   

\begin{figure}
\includegraphics[width=0.4\textwidth]{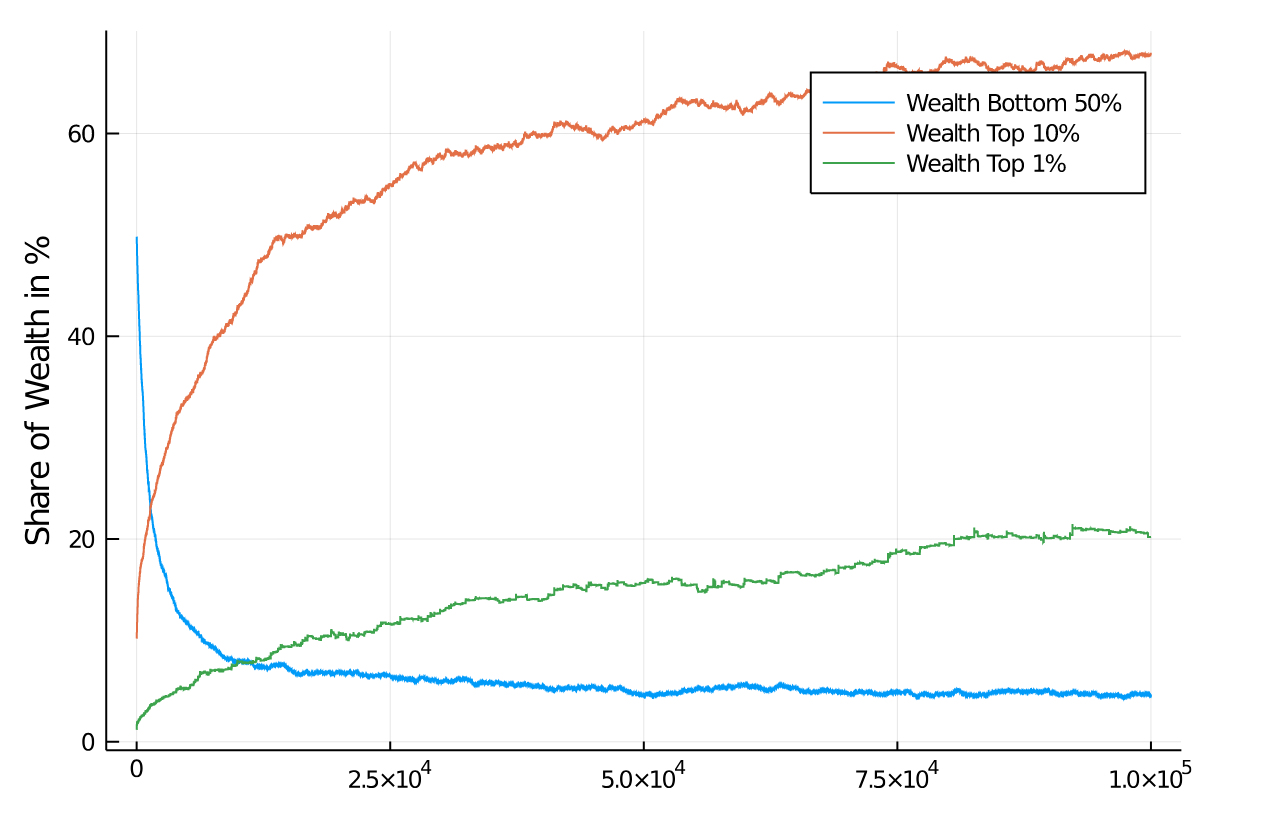}
\includegraphics[width=0.4\textwidth]{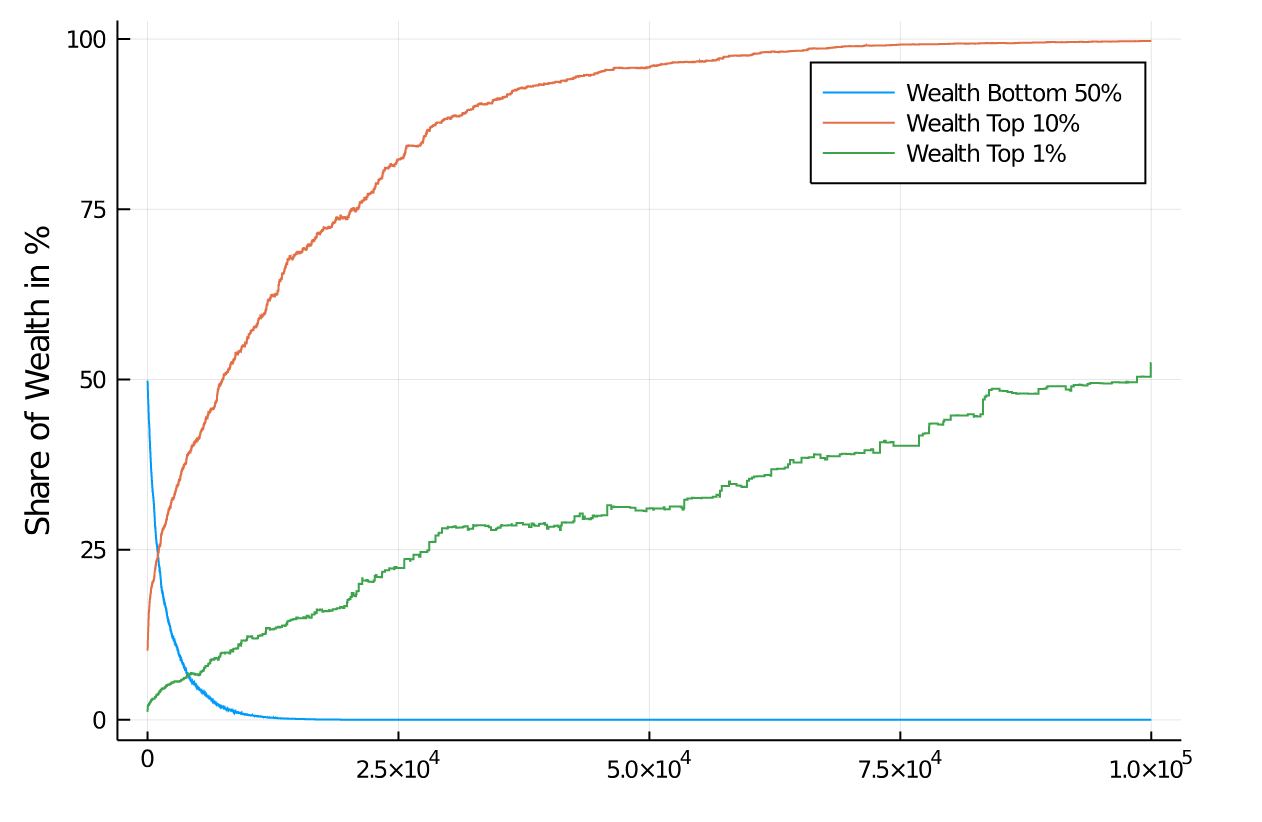}
\caption{Wealth development in different quantiles of the population of 1,000 agents. Flat income tax regime, with regular flat tax of 
30\%. Comparison of income tax application with redistribution to all, case I (above) and baseline case, {\it untaxed} (below). Initial distribution was equal among all 1,000 agents. Quantiles shown: Bottom half of the population vs top 10\% and top 1\% wealthiest agents.}	
\label{fig:quantiles}
\end{figure}

\begin{figure}
\includegraphics[width=0.4\textwidth]{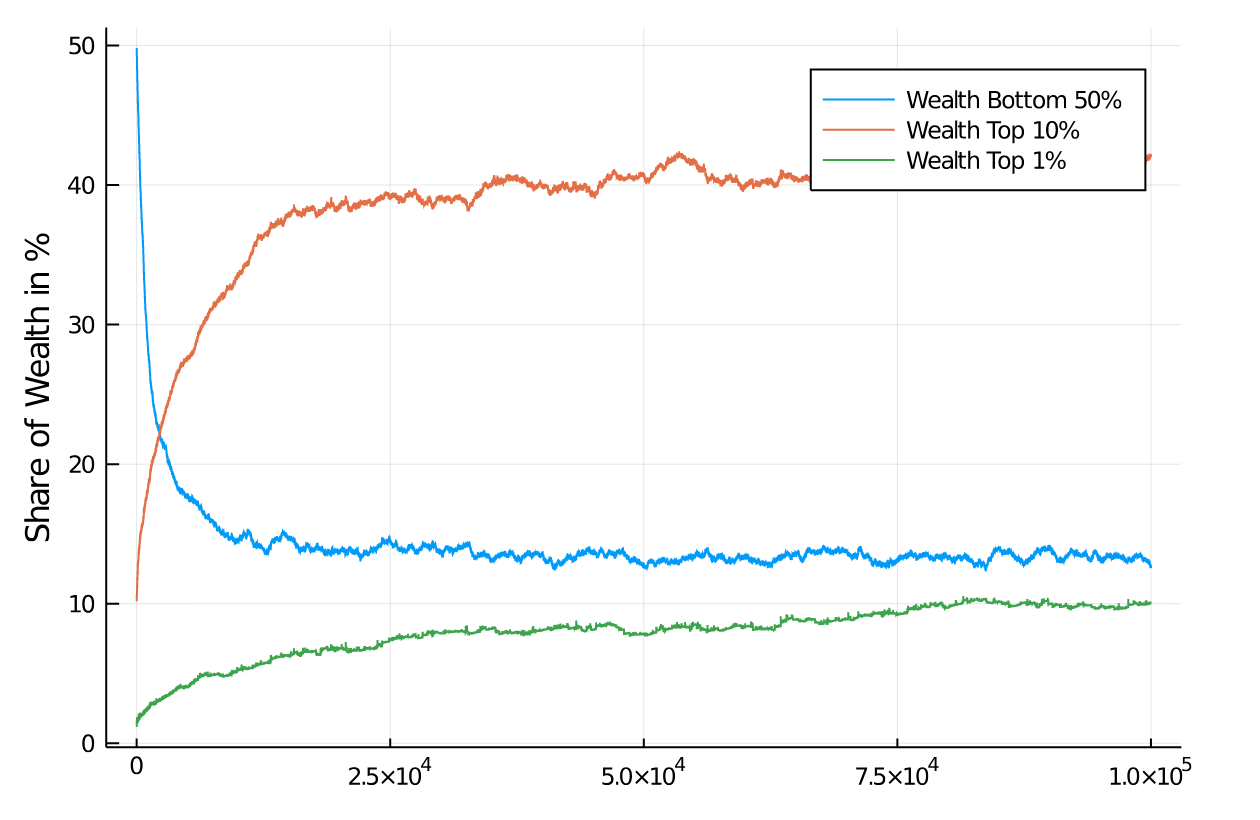}
\caption{Wealth development in different quantiles of the population of 1,000 agents. Flat income tax regime, with high flat tax of 
60\%. Initial distribution was equal among all 1,000 agents. Quantiles shown: Bottom half of the population vs top 10\% and top 1\% wealthiest agents.}	
\label{fig:quantiles-fit-high}
\end{figure}

One other way of looking at the distributional effects of taxes is to depict the development of wealth in certain quantiles of the population. Figure \ref{fig:quantiles} compares these distributional effects of the 30\% flat tax rate with the untaxed system.  So, there actually {\it is} an effect of a flat income tax policy on distribution of wealth (note the difference in scales) for different quantiles of the population, with the untaxed system approaching
100\% of wealth possession rather quickly, but this tendency is only dampened to a degree in the case of a flat income tax. A higher 
flat tax rate does further dampen the inequality, see Figure \ref{fig:quantiles-fit-high}, but even at 60\% flat tax rate very substantial inequality ensues. 

\begin{figure}
\includegraphics[width=0.4\textwidth]{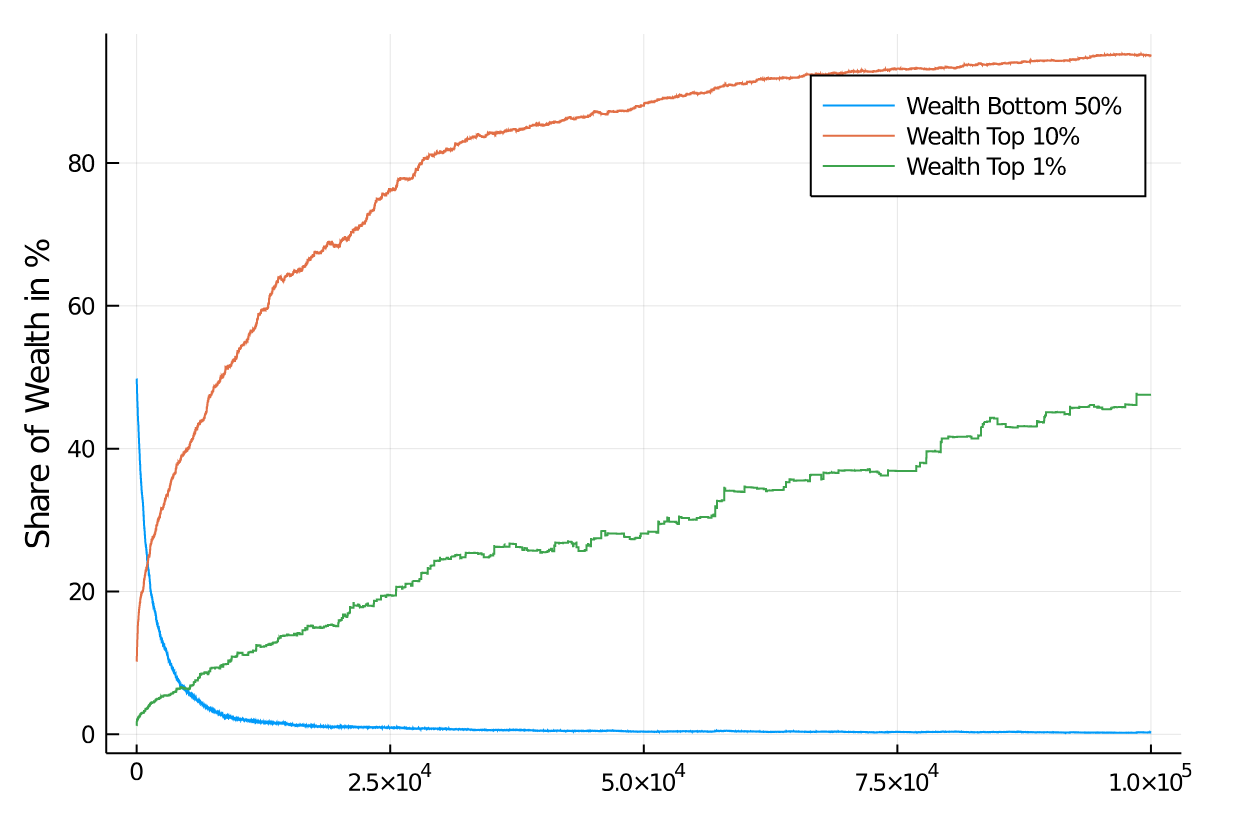}
\caption{Wealth development in different quantiles of the population of 1,000 agents. Flat income tax regime, with high flat tax of 
5\%. Initial distribution was equal among all 1,000 agents. Quantiles shown: Bottom half of the population vs top 10\% and top 1\% wealthiest agents.}	
\label{fig:quantiles-fit-low}
\end{figure}

We present a final set of simulations on a flat income tax by moving in the opposite direction: A low flat tax rate of 5\% (Figure  \ref{fig:quantiles-fit-low}) which again demonstrates the influence of  income tax rates on wealth inequality, though a weak one in this case.

\subsection{A Progressive Income Tax} \label{sec:fit-p}

Now we present simulations on a progressive income tax, applied with a rate of between 15\% and 45\%, 60\% and 75\%.
The lowest amount is applied for income above a certain tax-free threshold. While this is arbitrary, in light of the starting wealth of
each agent, we set the lowest rate to begin taxing at incomes of \$150. In order to keep things simple, we linearly increase the tax rate between
this minimum and the maximum being reached at \$850, \$1,200 and \$1,550, respectively, for the above mentioned maximal tax rate.
So there is some amount of income free of taxes, but the tax rate quickly rises to the maximum value.

\begin{figure}
\includegraphics[width=0.4\textwidth]{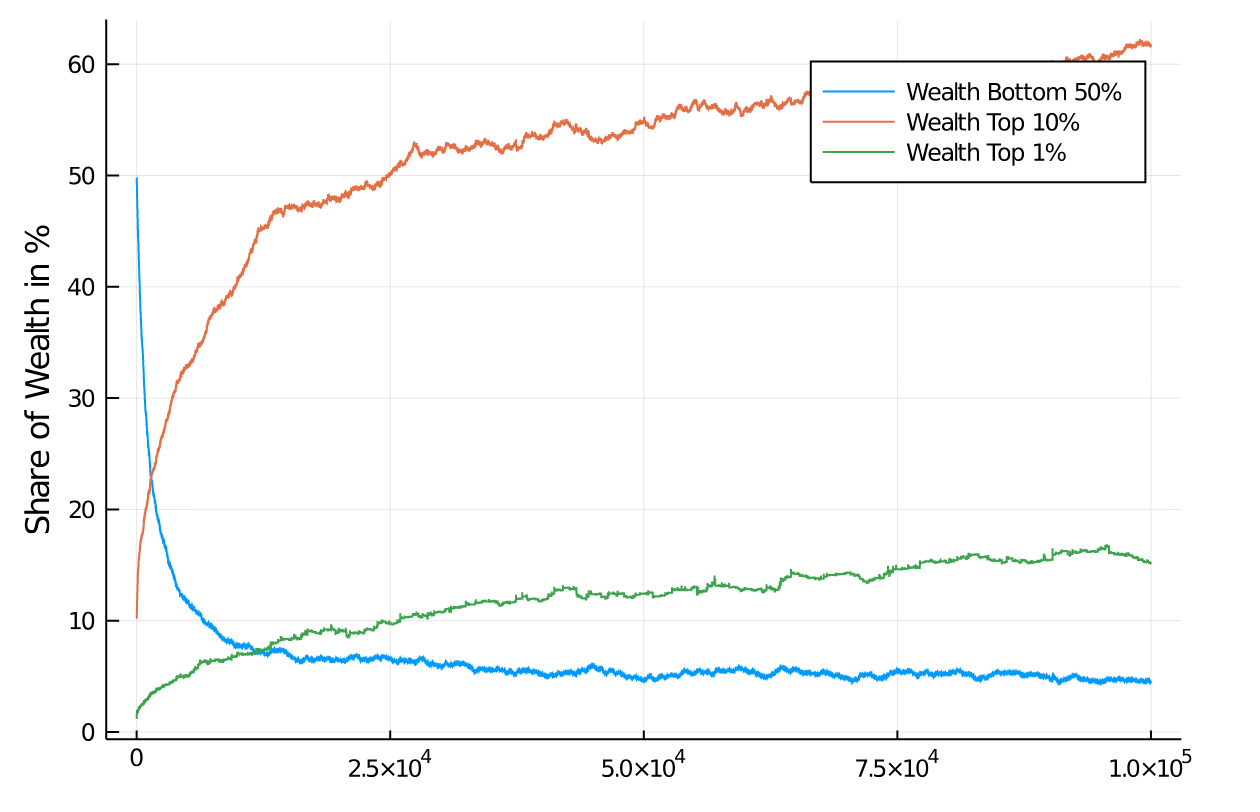}
\caption{Wealth development in different quantiles of the population of 1,000 agents. Progressive income tax regime, with tax rate of 
between 15\% and 45\%, linearly growing in the income interval from \$150 to \$850. Initial distribution was equal among all 1,000 agents. Quantiles shown: Bottom half of the population vs top 10\% and top 1\% wealthiest agents.}	
\label{fig:quantiles-pit-regular}
\end{figure}

\begin{figure}
\includegraphics[width=0.4\textwidth]{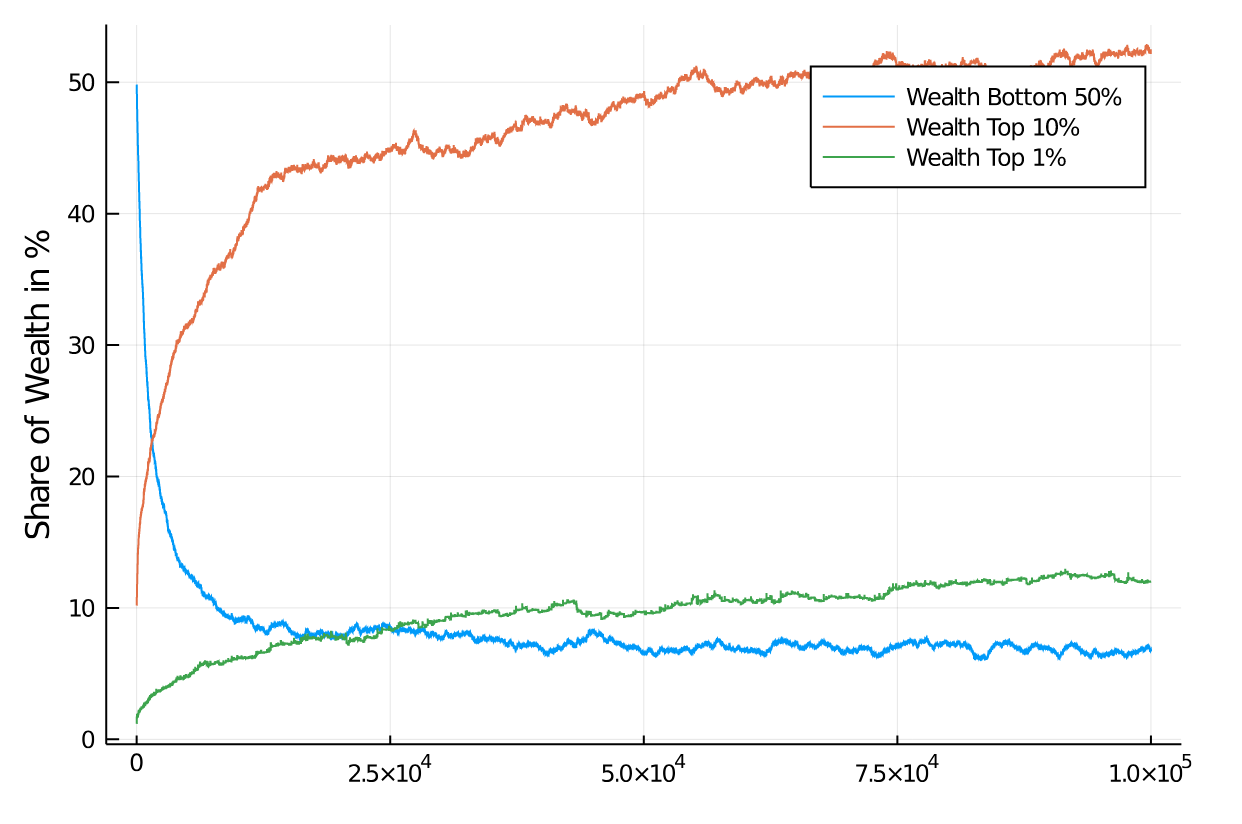}
\includegraphics[width=0.4\textwidth]{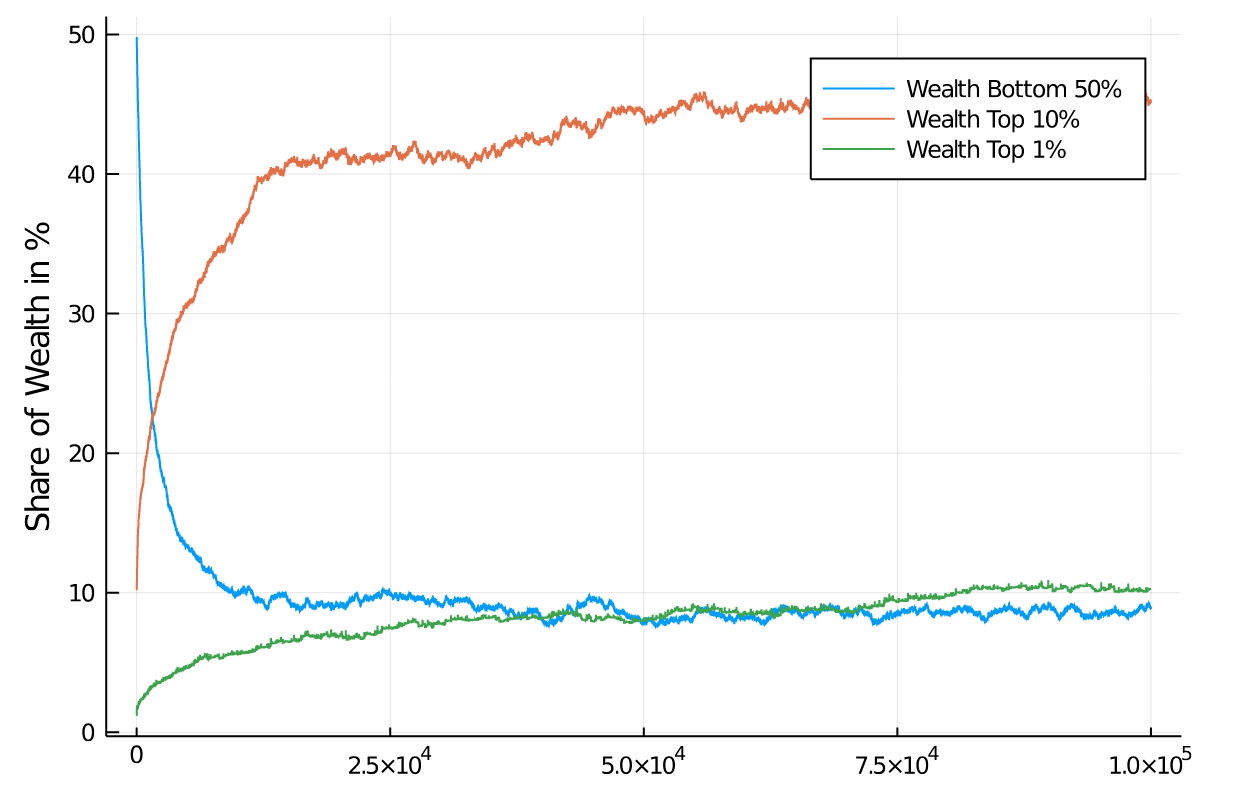}
\caption{Wealth development in different quantiles of the population of 1,000 agents. Progressive income tax regime, with tax rate of 
between 15\% and 60\%, linearly growing in the income interval from \$150 to \$1,200 (top) and between 15\% and 75\%, linearly growing in the income interval from \$150 to \$1,550 (bottom). Initial distribution was equal among all 1,000 agents. Quantiles shown: Bottom half of the population vs top 10\% and top 1\% wealthiest agents.}	
\label{fig:quantiles-pit-med-high}
\end{figure}

Figure \ref{fig:quantiles-pit-regular} shows the development of the distribution of wealth in the population over 100,000 iterations. The effects of this tax regime are quite similar to those of a flat tax of 30\%. Figure \ref{fig:quantiles-pit-med-high} (top) shows the application of a 60\% marginal tax rate, and Figure \ref{fig:quantiles-pit-med-high} (bottom) shows the application of a serious 75\% marginal tax rate. 

While it is obvious that there is an effect of the higher marginal tax on the development of wealth, the effect is not as serious as one might expect from a marginal tax rate of 75\%! In fact, the redistributive effect is quite disappointing, despite an effort that sounds
very serious. The top 1\% have gained a share of 10\% of the assets while the bottom half has lost most of its half of the assets and stands at about 8\% of assets after 100,000 iterations.

In summary, a very simple Artificial Chemistry model of economic activity that allows a starting state of all agents having equal wealth
-- a condition no society can hope to start from --  will always quickly develop economic inequality under an income tax regime intended to distribute wealth. We are not saying that income taxes do not have an effect, in fact, it can be seen that larger marginal tax rates do have effects, but these effects are tiny compared to the goal of keeping economic equality in a society. If one further adds the consideration that no society starts with perfect equality, but would have to start from a situation of economic inequality at the outset, 
the outlook for income tax remedies is even worse.

\subsection{A Wealth Tax} \label{sec:wt}

We now turn to another tax model that is not based on income, but on accumulated wealth. This tax is applied to the total of an agent's wealth, but the frequency of its application is reduced to a tenth. That is approximately the relation between applying a tax every month vs once a year. Here we are not concerned about the practicality of such a tax, but its mere ''theoretical'' application and effects.

\begin{figure}
\includegraphics[width=0.46\textwidth]{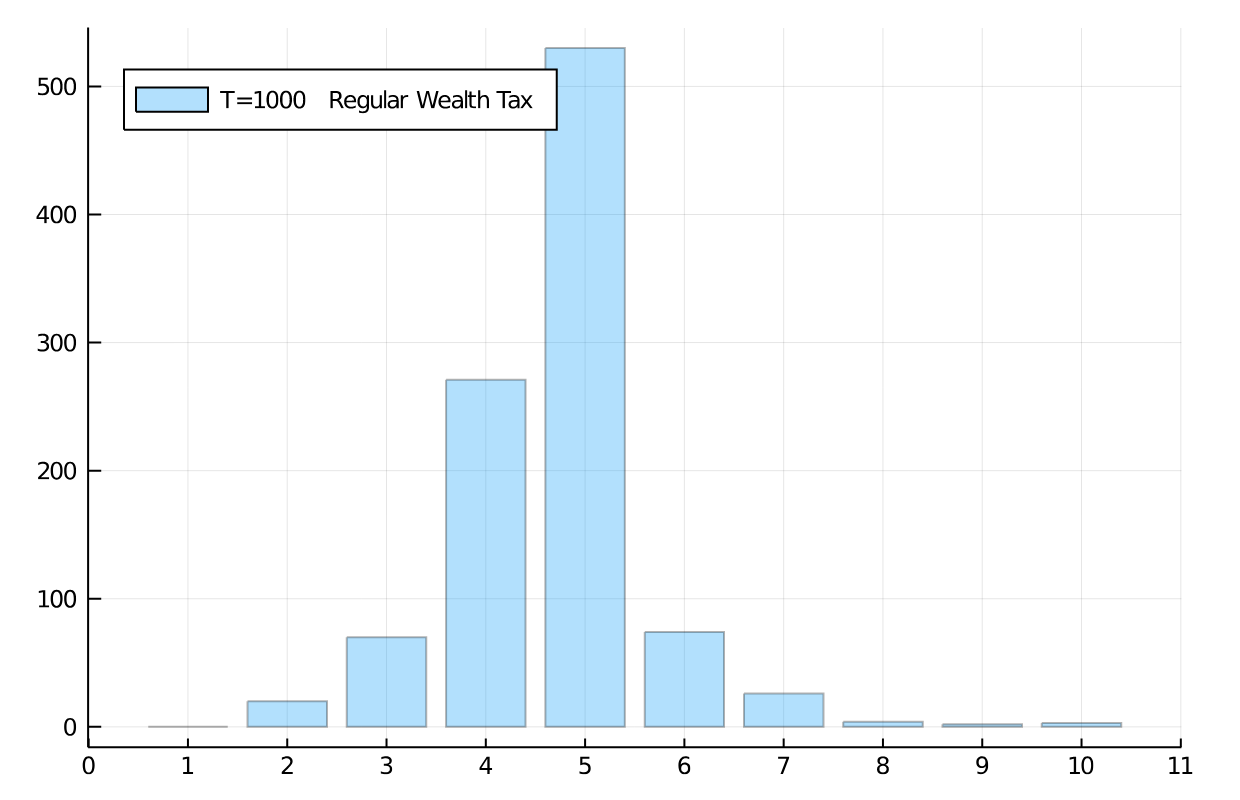}
\includegraphics[width=0.46\textwidth]{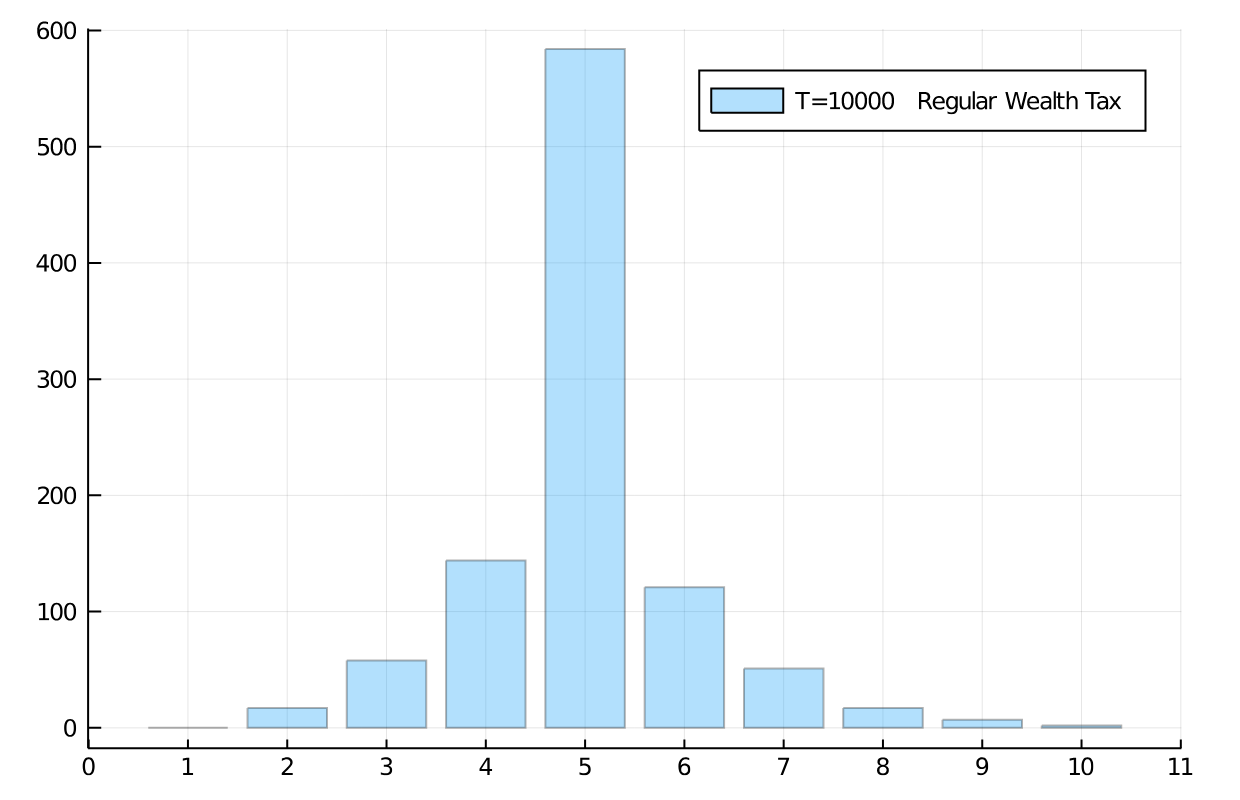}
\includegraphics[width=0.46\textwidth]{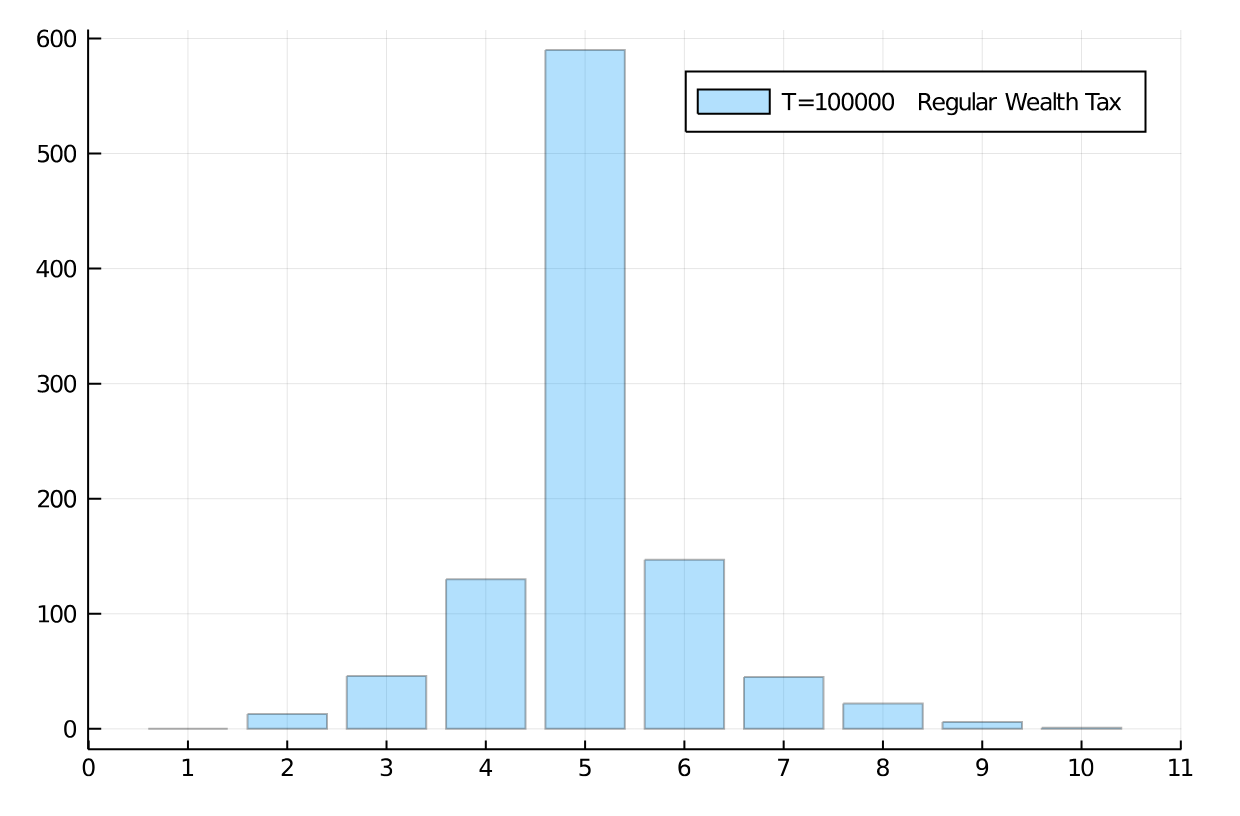}
\caption{Development of wealth distribution in population of 1,000 agents (10 bins, a.u.). Flat wealth tax regime, with regular flat tax of 
30\%  at $t=1,000; 10,000;$ and $100,000$ iterations. Initial distribution was equal among all 1,000 agents. Very effective redistribution of wealth, keeping the middle class dominant.}	
\label{fig:flat_wealth_tax-regular}
\end{figure}

Our first wealth tax, which simply takes each agent's wealth and subjects it to a flat tax we call regular is applying a rate of 30\%. Thus, at a frequency 10 times lower than the previous income tax which taxed the {\it difference} in wealth in a given period, we now tax the entire possession of the agent. Figure \ref{fig:flat_wealth_tax-regular} shows the distributional effect of such a wealth tax after $1,000; 10,000$ and $100,000$ iterations. The total of the tax is redistributed equally to all agents, our previous case I.

\begin{figure}
\includegraphics[width=0.4\textwidth]{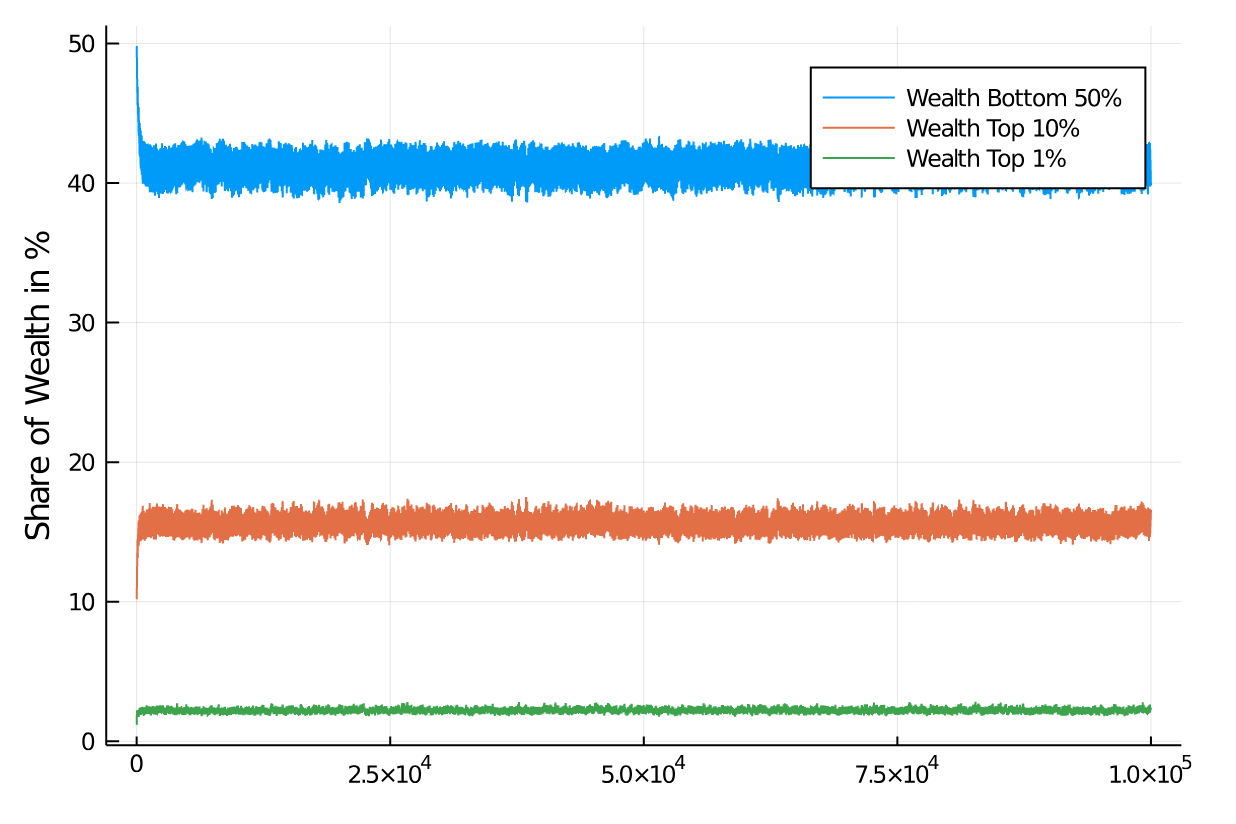}
\caption{Wealth development in different quantiles of the population of 1,000 agents. Initial distribution was equal among all 1,000 agents. Quantiles shown: Bottom half of the population vs top 10\% and top 1\% wealthiest agents.}	
\label{fig:quantiles-wt-regular}
\end{figure}

Figure  \ref{fig:quantiles-wt-regular} shows how the quantiles develop under such a tax. As is clearly visible, after an early relaxation phase, wealth distribution of certain quantiles of the population are quite stable in a band. The bottom half of the population quickly looses around 10\% of its share, but remains stable thereafter. The top 10\% of the population gain around 5-6\% of additional wealth
while the top 1\% moves up to a share of approximately 2.5\%. While these are big numbers still, they are in no way comparable to the quantile development with an income tax. The reason is that always the full amount of wealth accumulated by an agent is the basis of taxation. This, in combination with the systematic redistribution of all proceeds from the tax allows the agents to develop their wealth in only a tiny band. Figure \ref{fig:flat_wealth_tax-regular} actually shows a healthy distribution, with most agents in the middle bin (''middle class''), and a more or less symmetric, but quickly falling occupation of bins higher and lower in wealth.

\begin{figure}
\includegraphics[width=0.23\textwidth]{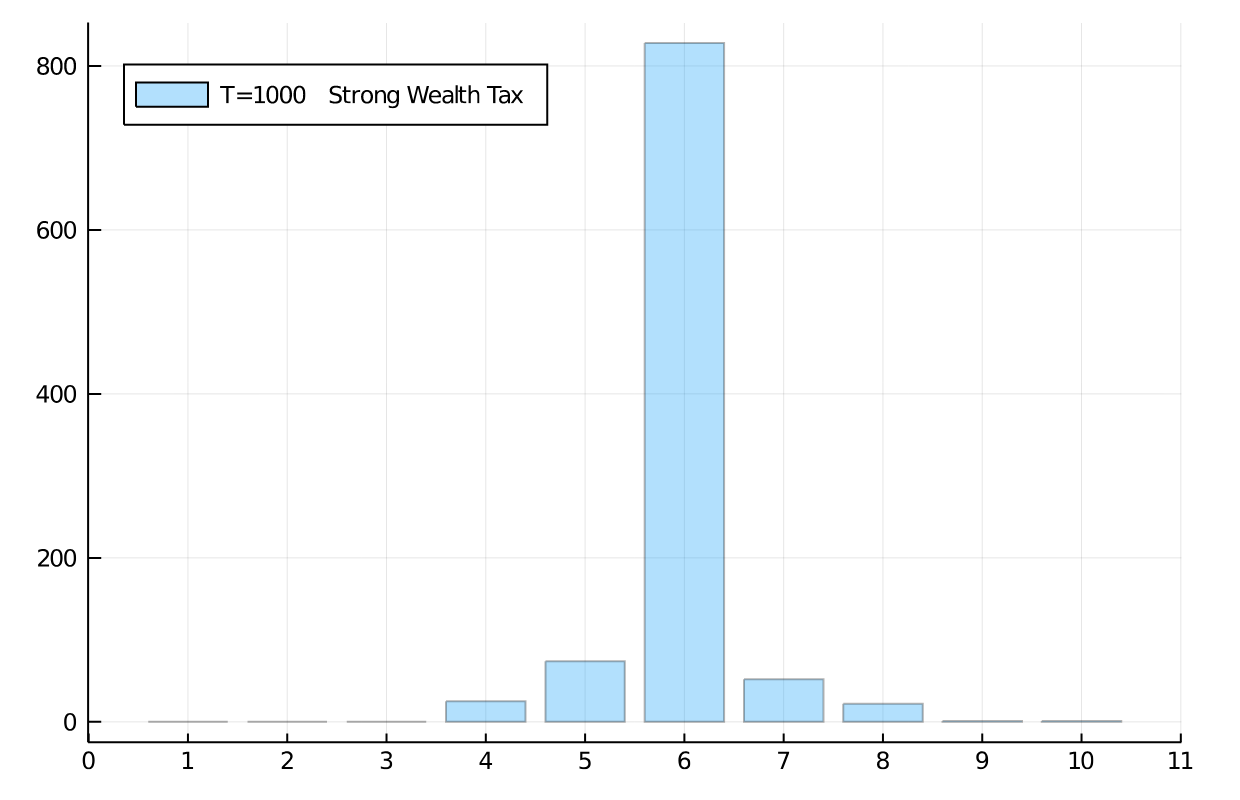}
\includegraphics[width=0.23\textwidth]{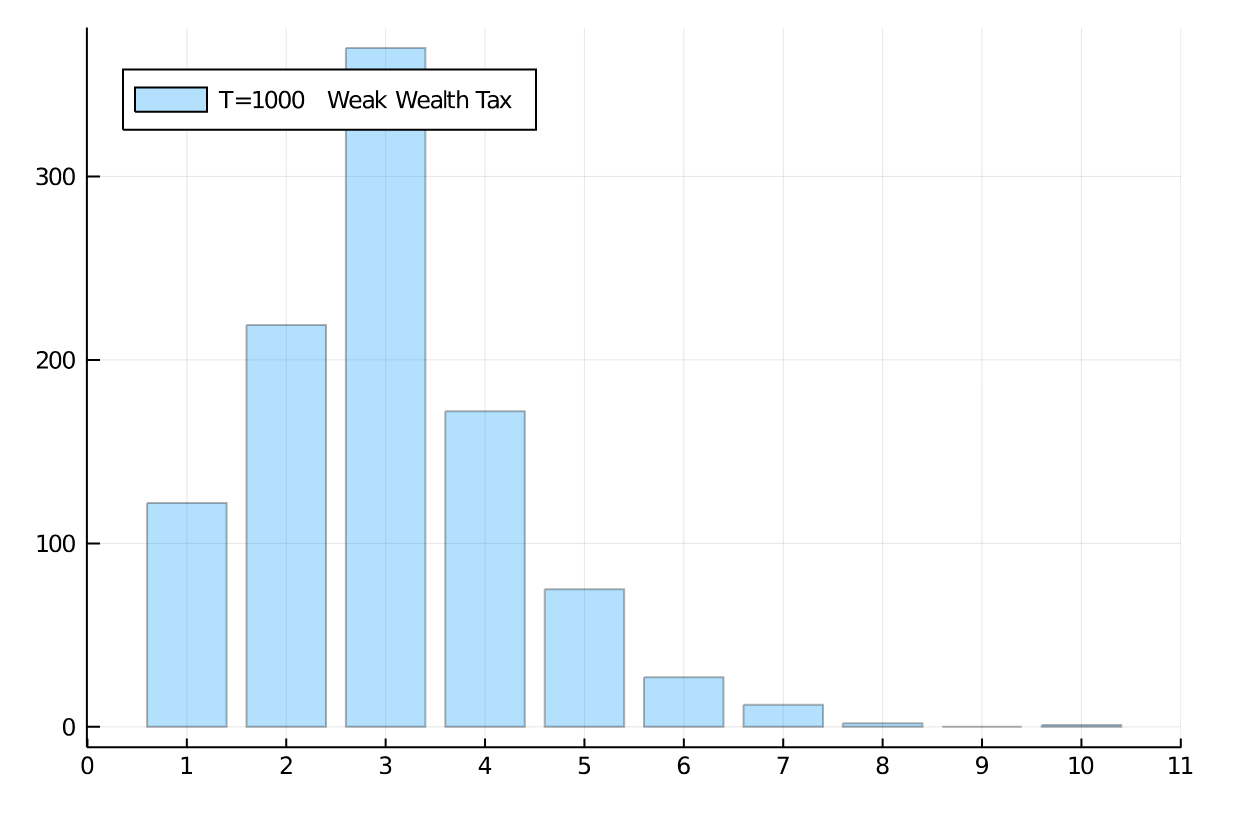}
\includegraphics[width=0.23\textwidth]{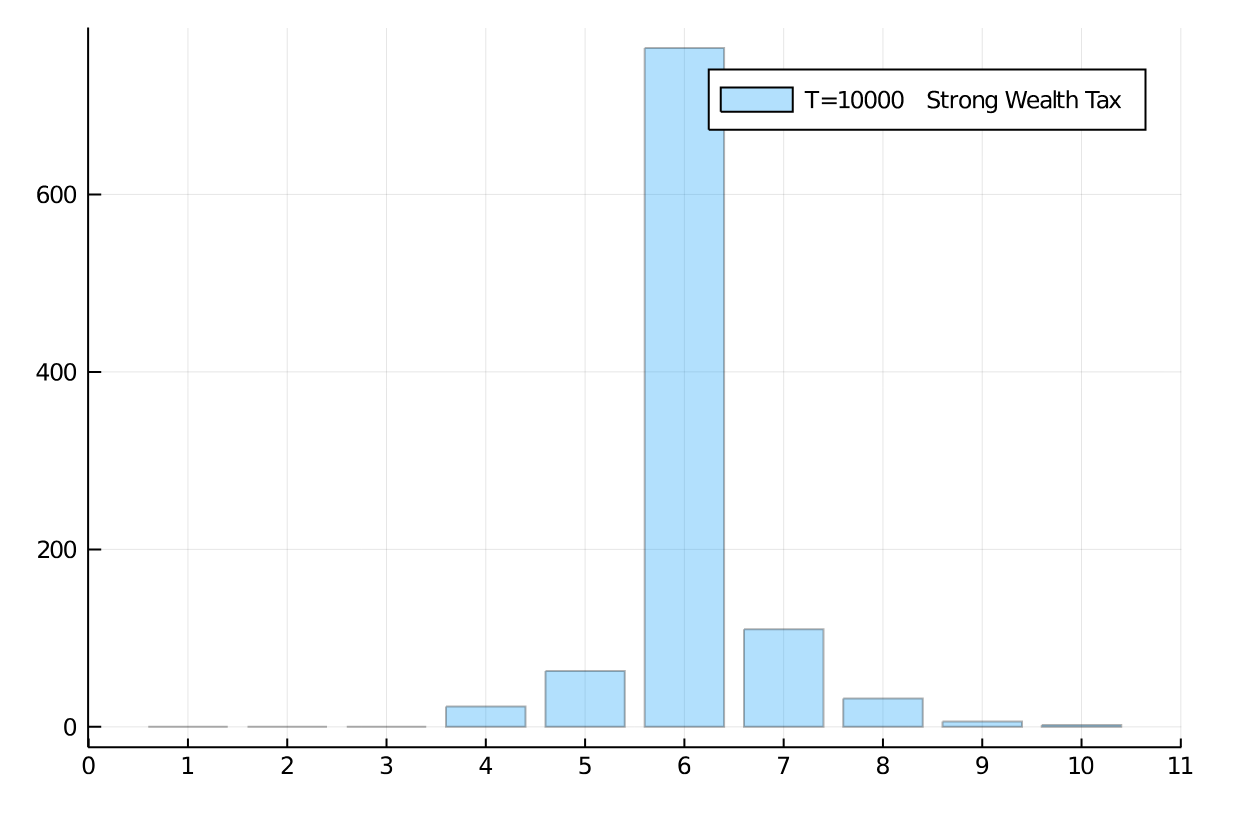}
\includegraphics[width=0.23\textwidth]{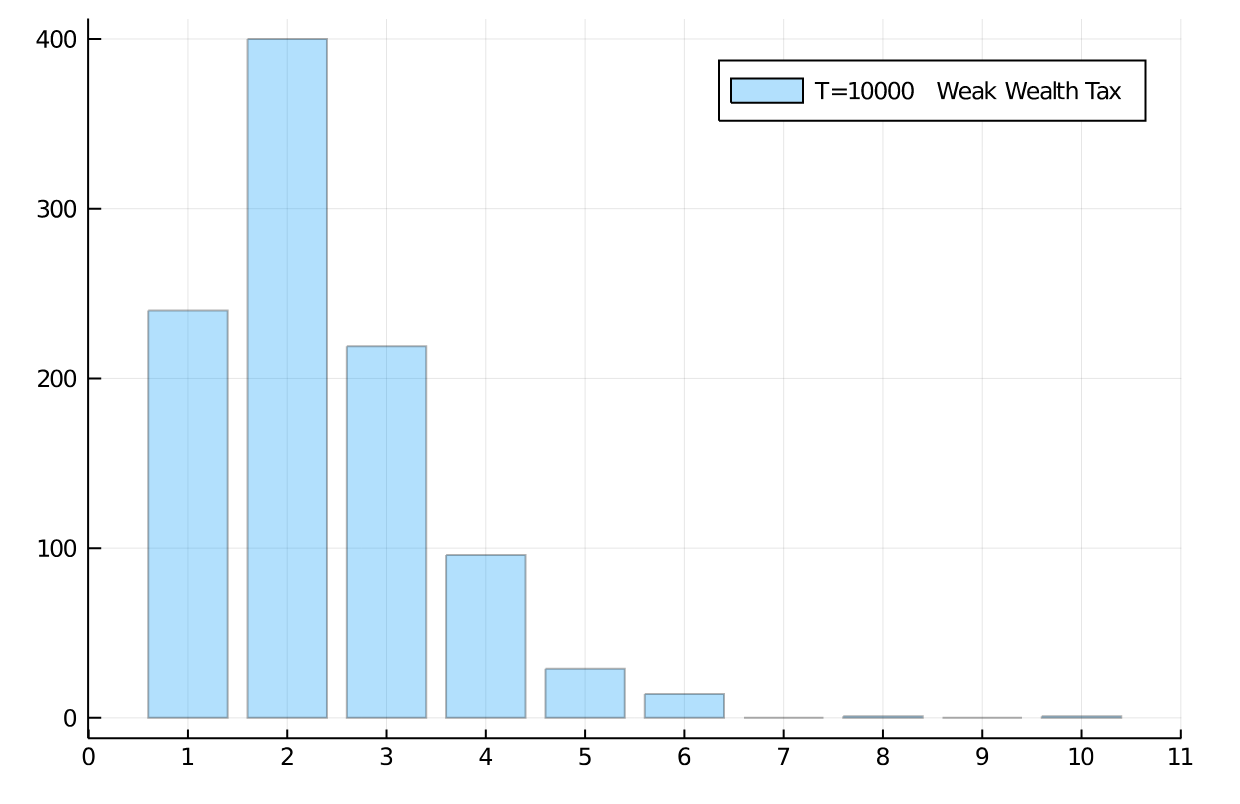}
\includegraphics[width=0.23\textwidth]{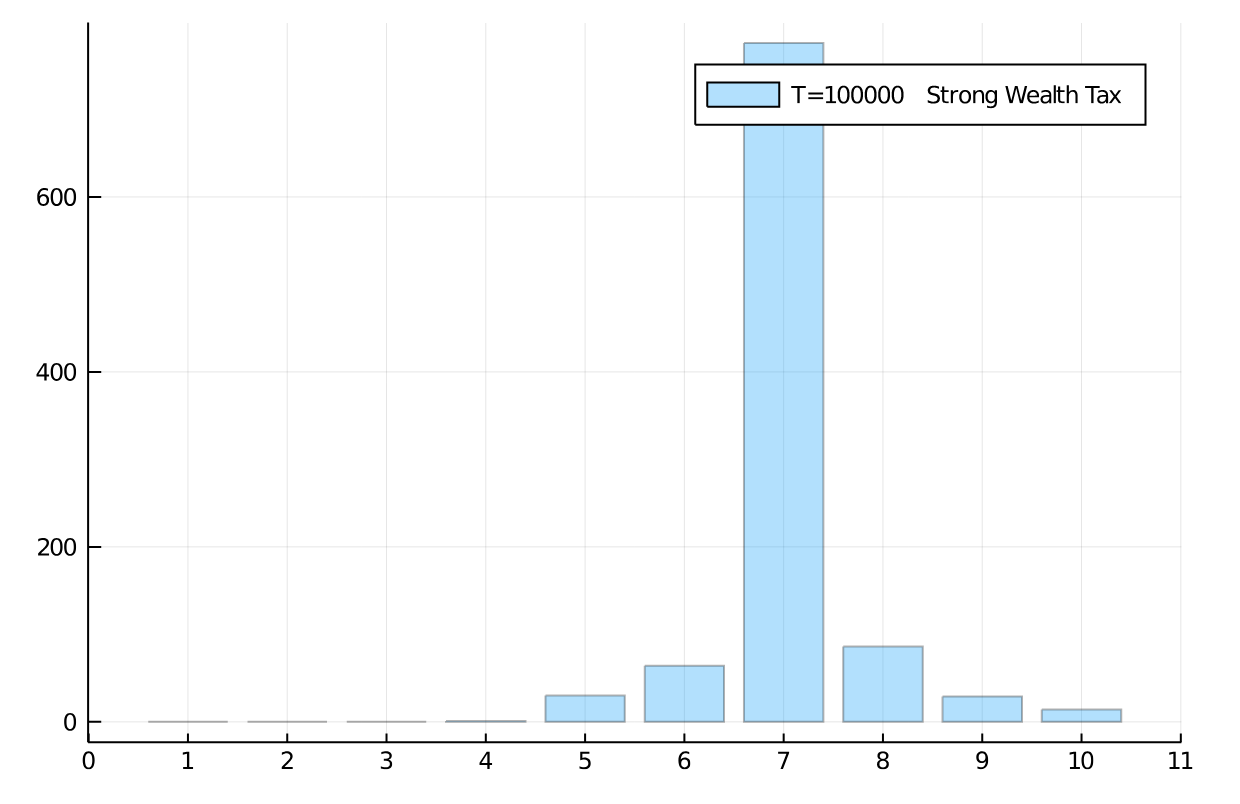}
\includegraphics[width=0.23\textwidth]{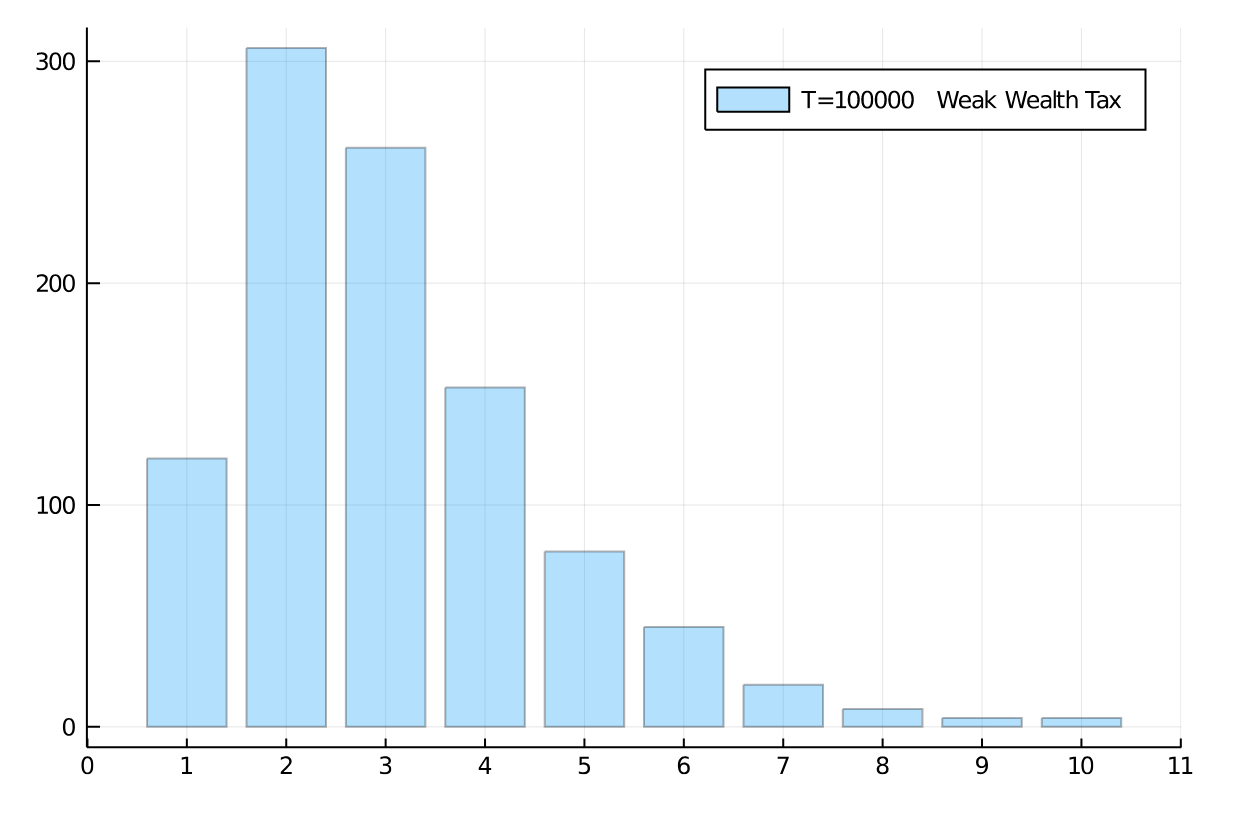}
\caption{Wealth distribution in population of 1,000 agents (10 bins, a.u.). Flat wealth tax regime, with high (60\%) and low (5\%) flat tax of 
30\%. Comparison of high (left) and low (right) for redistribution policy at $t=1,000; 10,000;$ and $100,000$. Initial distribution was equal among all 1,000 agents. High wealth tax increases wealth of middle class, low wealth tax reduced it in favour of a few high net-worth individuals.}	
\label{fig:flat_wealth_tax-high-low}
\end{figure}

\begin{figure}
\includegraphics[width=0.4\textwidth]{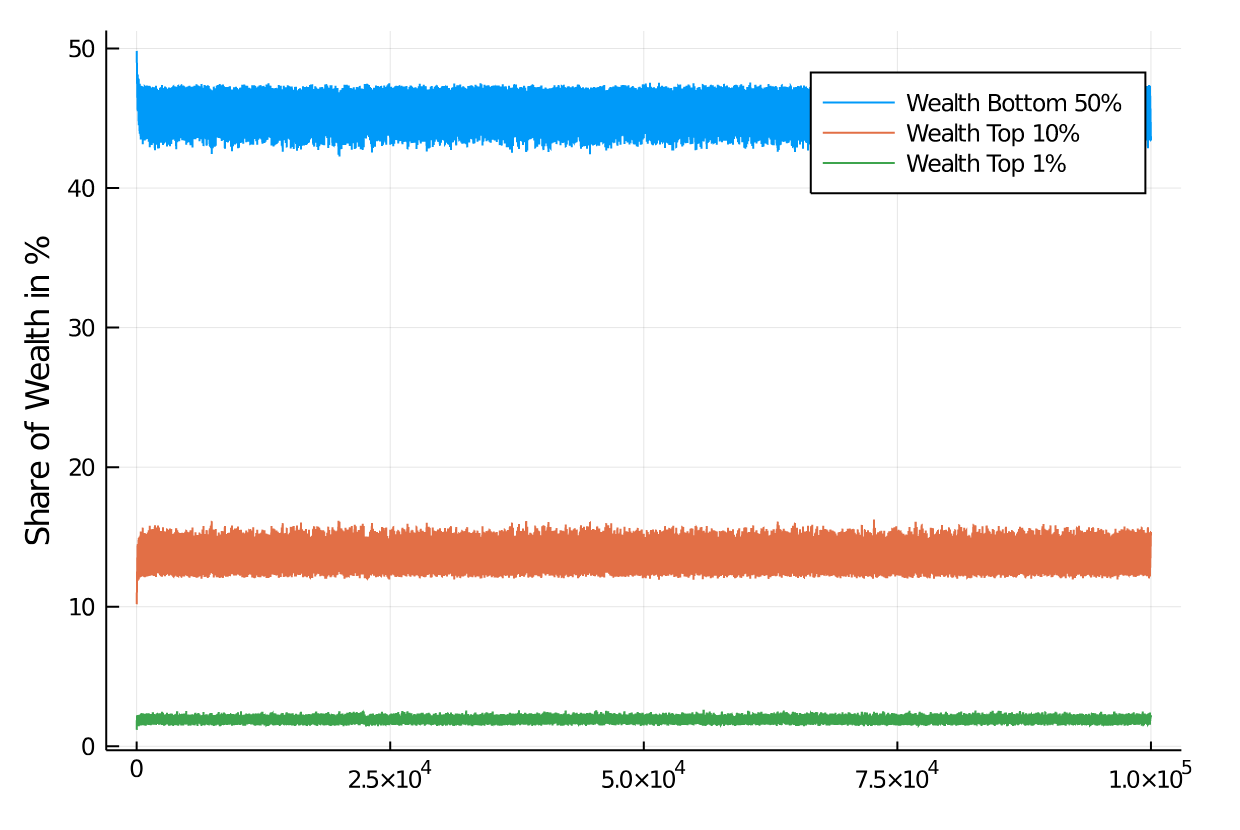}
\includegraphics[width=0.4\textwidth]{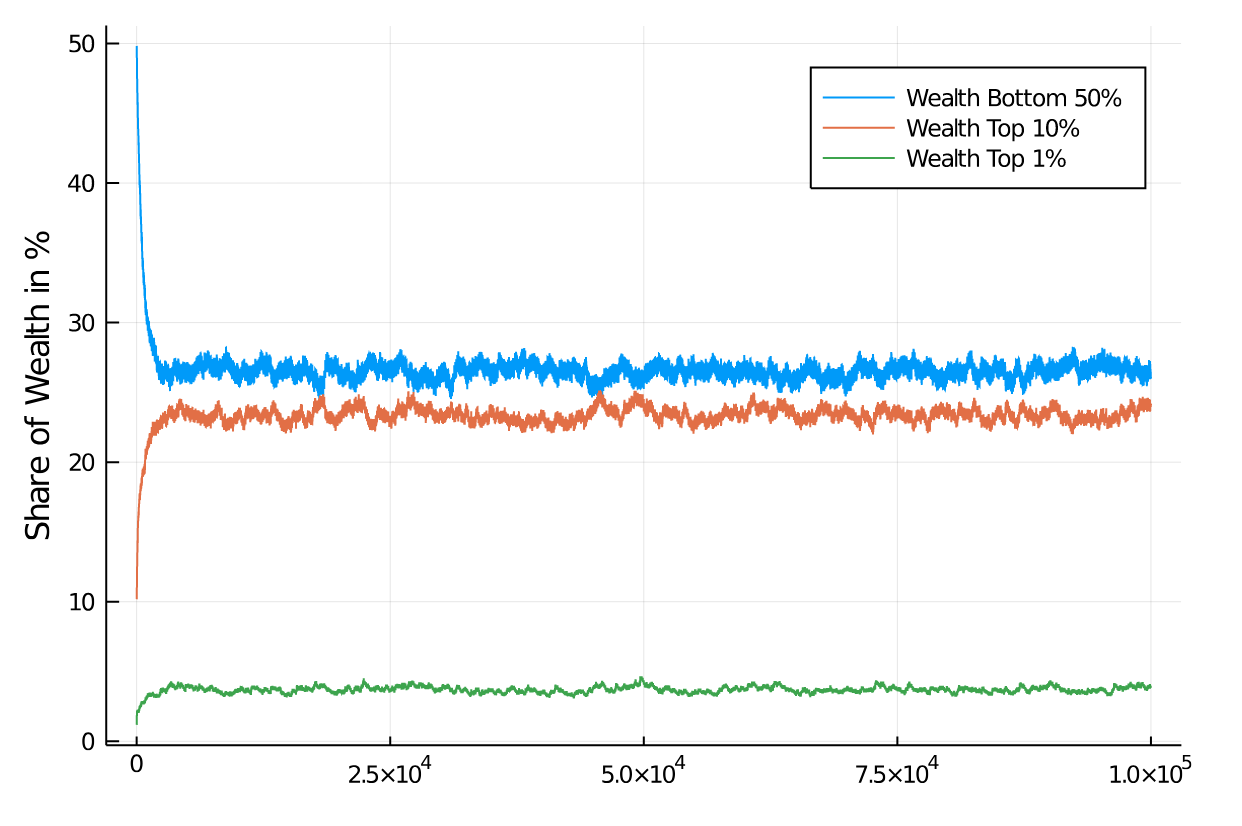}
\caption{Wealth development in different quantiles of the population of 1,000 agents. Different flat wealth tax regime, with tax rate of 
of  60\% (top) and 5\% (bottom). Initial distribution was equal among all 1,000 agents. Quantiles shown: Bottom half of the population vs top 10\% and top 1\% wealthiest agents. Distributions quickly stabilize after a short relaxation period.}	
\label{fig:quantiles-wt-high-low}
\end{figure}

We also examined a high wealth tax regime of taxing wealth at 60\% and a low regime of taxing it at 5\%, see Figures \ref{fig:flat_wealth_tax-high-low} and \ref{fig:quantiles-wt-high-low}. What is striking is that a wealth tax in general is able to 
quickly stabilize a distribution of wealth in the population, of course at different levels depending on the severity of the tax. We can see from the
graphs that a 60\% wealth tax does not add much distributional utility compared to only half of that rate (the regular case). A low wealth tax (here 5\%) as it is conventionally discussed in economic circles as a maximum measure (and normally discarded), on the other hand, leads to a more familiar distribution of wealth in the population. If we choose an even lower tax rate for such a flat wealth tax, say a 1\% rate, we can see effects similar to a strong income tax (see Appendix). 

These results seem at first sight somewhat counter-intuitive. Why would a wealth tax have so much stronger distributive effects than an income tax? After all, an income tax is supposed to tax the changes in wealth, so shouldn't it have the same effect as an admittedly smaller tax on the entire wealth? The answer is ''yes''. But one needs to keep in mind that income is only a tiny portion of overall wealth of an individual agent, and its influence in our model shrinks as economic inequality grows larger (since exchange is determined by 
the agent with smaller wealth). Thus, an income tax becomes progressively less effective in curbing the differential effects of wealth in a population. From our simulations, we can see that a wealth tax of approximately 1\% corresponds to a high flat income tax of 60\%, a close to two orders of magnitude difference in effectiveness! 

Note that the assumption of an asymmetric role of agents in an exchange is crucial. Absent the ability to go into debt, an agent can only afford and pay for goods and services up to the value of their wealth (but normally even to a smaller amount). Thus, the larger the difference in wealth between two agents, the smaller 
the amount of exchange in relation to the wealth of the agent with larger wealth, and therefore, the smaller the effect of their income tax.  
In other words, as economic inequality grows, so does the {\it inefficiency} of income taxes.

\section{DISCUSSION} \label{sec:discussion}

The natural question to ask, then, is how would one implement such a wealth tax on a larger scale and in reality? This question is especially important, given the counter-arguments that will be recruited to discourage any attempt at this. In the following we shall discuss these
aspects in particular in light of
\begin{itemize}
\item The attempts at tax avoidance
\item The availability of funds for a wealth tax (the liquidity problem)
\item The redistribution vs. government support role of taxes 
\item Issues of enforcement and practical measures
\item The possibility of a combinations of a wealth tax with a guaranteed basic income or other ideas 
\end{itemize}

Before we address these implementation questions, we would like to discuss some of the obvious reservations raised by opponents of a wealth tax. 
A non-comprehensive list of counter-arguments from economists could include the following arguments: 
\begin{itemize}
\item
The model builds a closed system, with no agents coming and going. --- This is a correct observation, but in contrast
to many other models of the economy, in principle, we can randomly introduce and remove agents from this system
 (akin to birth and death processes). This can be examined, but it will not bring substantially different other aspects 
 into consideration, except that one might want to study the effects of an inheritance tax. But the nice thing about a wealth is
 that an inheritance is, sooner or later, appearing as a contribution to wealth, and already covered by a wealth tax, if only with a delay.
\item
Another argument, frequently brought up against kinetic exchange models, and transferable to this model is that production 
is actually the main driver of an economy, not the exchange between participants. Any production or gain in productivity
would not be reflected here. --- Again, this observation is correct, but this is not the point of the model. In fact, absolute values
of wealth are not studied here, we instead only look at the relative position of agents in the wealth hierarchy. For the study 
of distribution, which is relative among agents, the influence by the absolute growth of the economy is not of interest. 
We could even add an amount akin to production gains to the amounts to be distributed, but that will not make distributive
changes. 
\item
The model does not accurately reflect the wealth distribution found empirically. --- Again, this is by-and-large correct, but not 
the point of this model either. This model studies the rough distributional effects of taxation, not the exact outcome. There are
other models that attempt much closer accuracy when modelling, see, e.g., savings efforts and their effects on the long tail of the distribution in kinetic exchange models \cite{chakra2013}. 
\item
All of this is well known since the 1960s. --- Perhaps yes, but given the critical importance of a wealth tax as a remedy against economic
inequality, it seems that it is not worked on properly for serious implementation purposes. We thus would argue for more research and a closer look at relatively high wealth taxes and their implementation.  
\item
Wealth in general is difficult to measure, and therefore it is difficult to tax. --- As a general observation, this is certainly true. But property taxes (on private homes) are an example of how wealth (in a certain branch of the economy) can be measured. The message is again one of relative versus absolute measurements. As long as the same principles are applied to wealth of the {\it same type}, their relative differences will go into the tax calculation. That is enough to have distributive effects. The actual weight of a type of wealth and the rate of redistribution is then subject to 
political considerations. 
\item
The real purpose of taxes is not redistribution, but state income. --- The aspect of state funding is certainly in need to be addressed. This can be done by removing part of the redistribution amounts and transferring them to the state. It is again an absolute, not relative amount we are talking about here, and as such, it cannot have distributive effects. 
\end{itemize}

Let us return now to the implementation questions posed above. 
\subsubsection*{Attempts at tax avoidance} Any tax system will have to deal with a certain degree of tax cheating and systematic tax avoidance schemes. As for the cheating, one cannot avoid that, but it will probably remain in the same proportion as our current tax systems. If anything, a tax system that is perceived as more just will likely be at the lower end of the cheating proportion. As for systematic tax avoidance, this can only be addressed by a certain degree of auditing. A tax system based on a wealth tax is not fundamentally different in this regard from a tax system based on income taxes. If anything, a simple flat wealth tax will probably 
be easier to administer and audit (in particular if there can be specialist audit teams formed for different types of wealth, that work together to determine overall amounts). As we indicated earlier, a likely candidate for examining avoidance effects is game theory.

\subsubsection*{Liquidity} This is indeed a problem for high net-worth individuals that might be taxed substantial amounts without being able to mobilize the necessary liquidity from the assets taxed. It is likely best to keep an income tax in place, and implement the calculation based on a wealth tax with the annual tax declaration which then would consider the income tax as an advance payment on the final wealth tax amounts. To be clear: We envision a tax system entirely based on a wealth tax, but contemporaneously relying on an income tax as a source of liquidity. This would include dividend and interest payments taxed at source (a withholding tax). Illiquid assets,  however, would likely have to be sold, at least partially to procure enough liquidity for tax payments. A pragmatic delay for required payments could be installed, also taking account of the amounts redistributed to everyone from the overall tax revenue. As for the amounts returned to taxpayers, these could be based on an estimate of the tax volume expected in any given year, and corrected
in a subsequent year with the actual amounts.

\subsubsection*{Government Support versus Redistribution}
Federal, state and local governments draw income from taxes and fees, but taxes are their primary source for providing services.  We have not modelled such a purpose of taxes in this contribution. However, it can be easily accommodated if the system is opened to the outside world and production is introduced as a source of additional income. Our assumption then would be that the government's share of income cannot be larger (and should preferably be smaller) than production income.

\subsubsection*{Issues of Enforcement and Practical Measures}
Enforcing the payment of a tax is best done by collecting it at source. This cannot easily be done with a wealth tax, so it is better to tax every income at an appropriate rate, and then to compensate the year after. Certain countries are better equipped for enforcing tax payments, especially given the potential that wealth moves to another country, but the general approach would be a collaboration between nations. 

\subsubsection*{Combinations of a Wealth Tax with a Guaranteed Basic Income or other Ideas}
We do not exclude the possibility that the best course of action would be a combination of the wealth tax with some sort of guaranteed basic income. However, for now we have not included that or other modifications into our taxation scheme.

\section{CONCLUSION} \label{sec:conclusion}

The simulations reported here use a very simple model of economic activity. Basing a study of economic activity on artificial chemistries opens a different way of examining collective effects in economic models. Notably, we are completely free to define the interaction rules between agents based on what we believe is important and ignoring supposedly unimportant features, which includes breaking symmetries or conservation laws or other principles that play a role in other disciplines.  

The results shine a harsh light on the idea that fiddling with income tax systems can rectify the highly unequal distribution of economic assets that exists today in most societies. Income taxes are by definition only applied to {\it changes} in wealth and normally vanish for very small or negative incomes. They, therefore, cannot correct a situation that is unequal from the outset, at least not without a substantial redistribution {\it beyond} the revenue an income tax can generate, something like a stable basic income or a large basic personal deduction that can be monetized if not taken in as agent income. In our system this was demonstrated by the fact that income taxes were applied only to agents that engaged in economic activity.    

It is worth mentioning that the unequal distribution of wealth today is actually a good starting point for introducing a wealth tax. The reason is that effective tax rates are close to zero for most agents, and nonlinearly increase to both sides of the wealth distribution. That is due to the redistribution of revenue from this tax as it was introduced here. In a situation where most agents are at the lower end of the
wealth spectrum a comparatively even smaller percentage of agents will have to pay substantial amounts of tax. It can be safely assumed, that the scenario also allows for more mobility between segments of the society (at least as far as wealth is concerned). 

Real economies are perhaps positioned between the asymmetric and symmetric cases of economic activity discussed here, 
with the symmetric case (kinetic exchange) more benign by having less pressure toward inequality than the asymmetric case used as 
a baseline in this manuscript. 
This would allow for some flexibility in regard to the actual rate of a wealth tax. However, if societies want to address their natural tendency to create inequality and avoid harmful ways of redistribution of unequal wealth, a wealth tax looks to be the single most 
effective tool to achieve progress toward a more just distribution.

\vspace{3mm}
\begin{acknowledgements}
W.B. gratefully acknowledges support from the John R. Koza endowment fund to Michigan State University. Computer simulations 
were executed with Julia v1.0.2 under MacOS 10.15.5. 
\end{acknowledgements}


\bibliographystyle{spmpsci-modi}
\bibliography{Inequality}

\pagebreak

\section*{Appendix}

\begin{figure}[!hbp]
\includegraphics[width=0.4\textwidth]{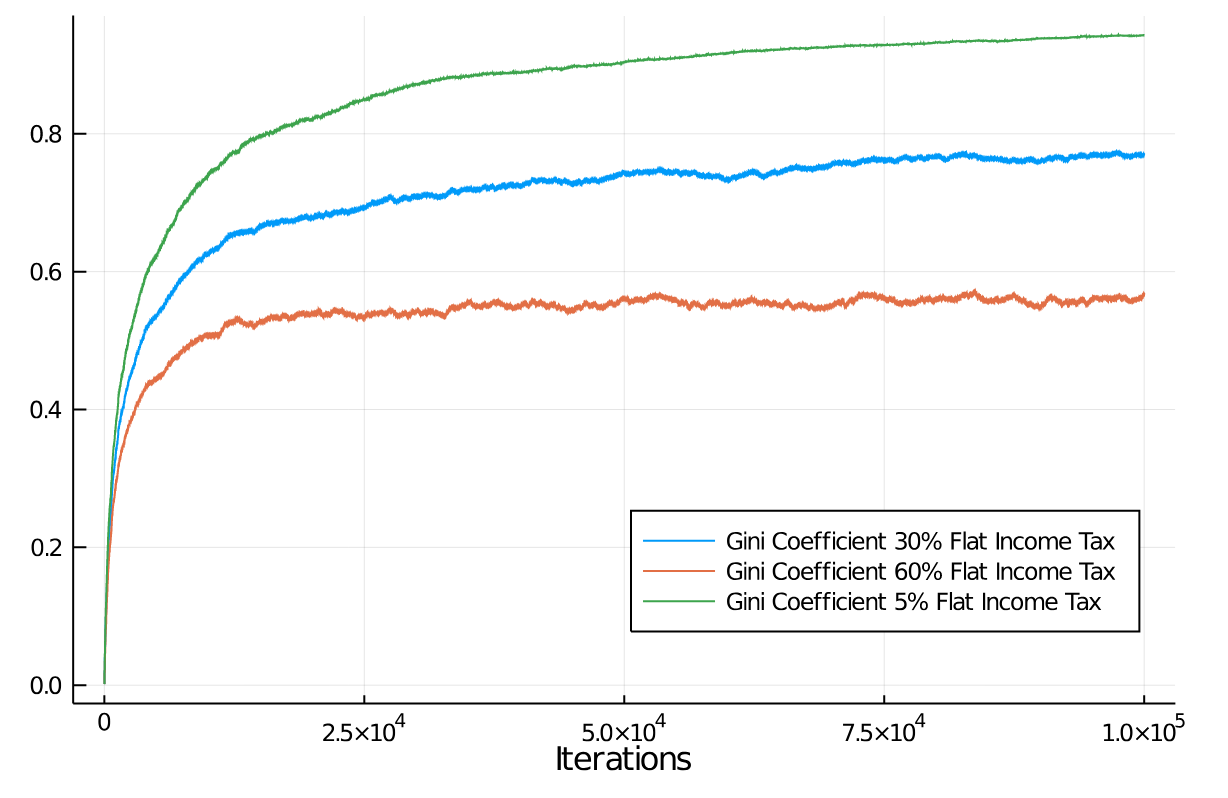}
\caption{Gini coefficient for different flat income tax regimes at 30\%, 60\% and 5\% with a population of 1,000 agents, starting from 
an equal distribution of wealth. Starting at ''0'', Gini indicates substantial inequality in all cases.}	
\label{fig:gini-fit}
\end{figure}

\begin{figure}[!hbp]
\includegraphics[width=0.4\textwidth]{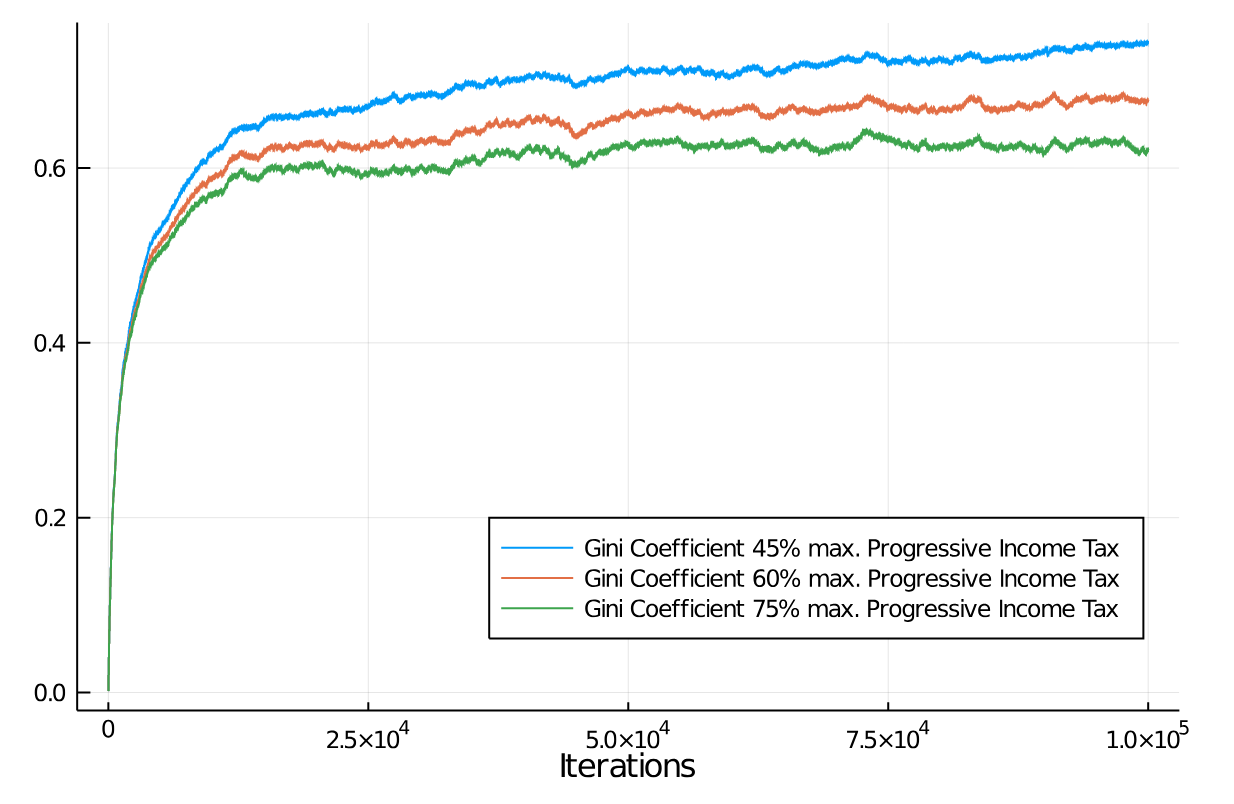}
\caption{Gini coefficient for different progressive income tax regimes with marginal tax rate of 45\%, 60\% and 75\% with a population of 1,000 agents, starting from an equal distribution of wealth. Gini indicates even more substantial inequality than flat tax rate in all cases.}\label{fig:gini-fit-p}
\end{figure}

Figure \ref{fig:gini-fit} shows the Gini coefficients for the flat income tax regime of Section \ref{sec:fit}, while Figure \ref{fig:gini-fit-p}
shows the Gini for a progressive income tax scenario (see Section \ref{sec:fit-p}).

As is clearly visible from the application of income tax schemes at different levels, their effect on wealth distribution is much smaller than the natural pressure for unequal distribution of wealth from normal economic activity. What is striking is that a progressive income tax in the current model does not even have the same effect as a flat tax, likely due to the small spread in income distribution through individual economic transactions in our model.


\begin{figure}[!htbp]
\includegraphics[width=0.4\textwidth]{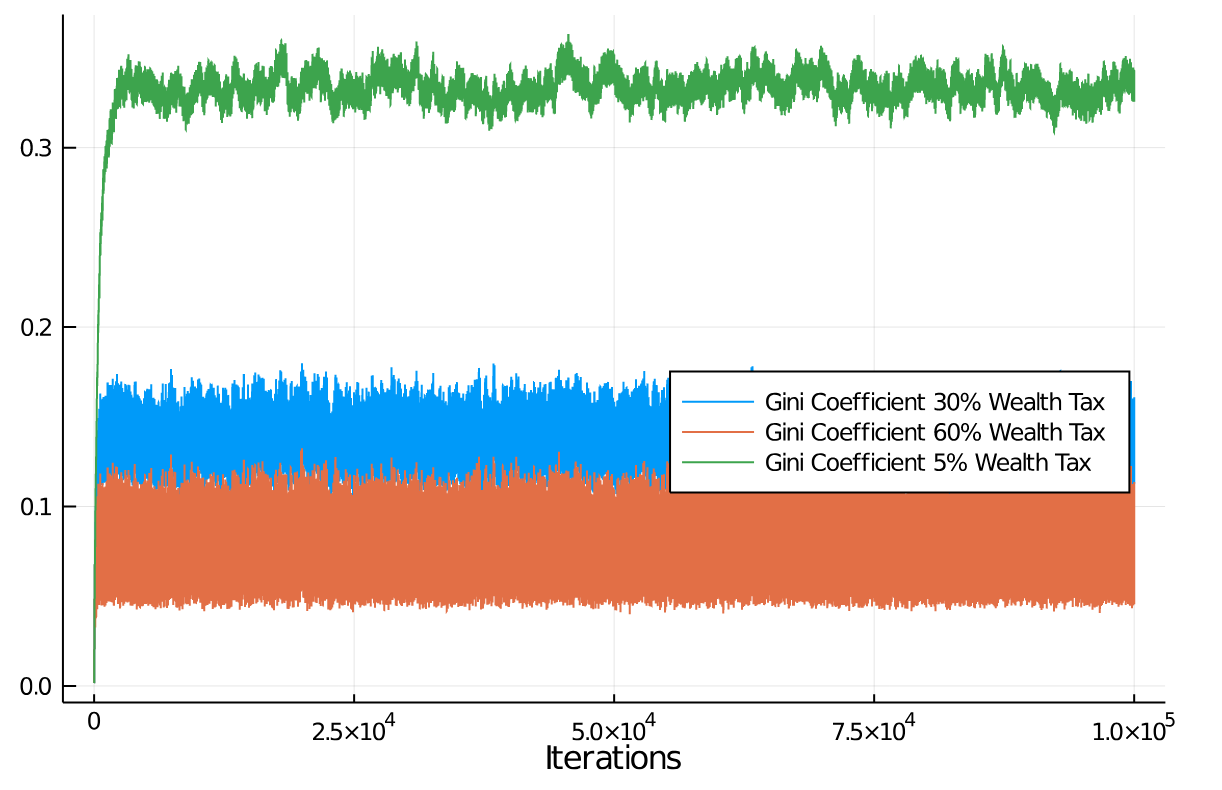}
\caption{Gini coefficient for different wealth tax regimes with a tax rate of 30\%, 60\% and 5\% with a population of 1,000 agents, starting from an equal distribution of wealth. Gini indicates good economic equality, with fluctuations in bands of 0.01 to 0.05, with a maximum value of the Gini coefficient of $0.35$ calculated for the lowest wealth tax rate.}
\label{fig:gini-wt}
\end{figure}

Turning finally to the effects of wealth taxes on economic inequality, we show Gini coefficients for the three wealth regimes we have considered in Section \ref{sec:wt}. Figure \ref{fig:gini-wt} shows the Gini for the three rates of a flat wealth tax we considered in Section \ref{sec:wt}. After quickly growing from ideal equality at ''0'', the Gini remains in a narrow band for all wealth tax regimes with an average of around 0.35 for the lowest wealth tax. If one were to apply a 1\% wealth tax only, though, the resulting wealth distribution would develop the classical exponential distribution, on par approximately with a 60\% flat income tax.

\end{document}